\documentclass[9pt,twocolumn]{extarticle}

\usepackage{soul}
\usepackage{titlesec} 
\usepackage[T1]{fontenc}
\usepackage{mathpazo} 
\usepackage[backend=bibtex,style=nature]{biblatex}

\DeclareNameAlias{sortname}{first-last} 
\DeclareNameAlias{default}{first-last}
\bibliography{references_schmidt_number.bib} 

\usepackage{dsfont}
\usepackage{graphicx}
\usepackage{subcaption}
\usepackage[margin=0.9in]{geometry}   
\usepackage[usenames,dvipsnames]{color}
\usepackage[colorlinks,linkcolor=Blue,urlcolor=Blue,citecolor=Blue]{hyperref}
\usepackage{amsmath,amssymb,braket}
\usepackage{amsfonts}
\usepackage{bbm}
\usepackage[squaren]{SIunits}
\usepackage{pifont}

\usepackage{array}
\newcolumntype{P}[1]{>{\arraybackslash}p{#1}}
\newcolumntype{Q}[1]{>{\centering\arraybackslash}p{#1}}

\usepackage{diagbox}

\usepackage{mathpazo}

\usepackage{courier}
\normalfont

\usepackage{physics}

\usepackage{latexsym}

\usepackage[T1]{fontenc}
\usepackage{caption}  
\usepackage{upgreek}

\newcommand{\ad}[1]{\textsuperscript{#1}\kern-2pt}

\usepackage{amsmath,amsthm,mathtools}

\makeatletter
\def\blx@maxline{77}
\makeatother

\usepackage{capt-of}

 \usepackage{flushend}

\usepackage[ruled]{algorithm2e} 
\usepackage{newfloat,algcompatible} 
\def\mytitle{Quantum stochastic communication via high-dimensional entanglement
\vspace{-2mm}}      
 
\title{\vspace{-1.cm}\huge\textbf{\textrm{\mytitle}}}

\author{Chao Zhang$^{1,2,7}$, Jia-Le Miao$^{1,2,7}$, Xiao-Min Hu$^{1,2,3\star}$, Jef Pauwels$^{4,5\star}$, Yu Guo$^{1,2}$, Chuan-Feng Li$^{1,2,3}$,\\ Guang-Can Guo$^{1,2,3}$, Armin Tavakoli$^{6\star}$, Bi-Heng Liu$^{1,2,3\star}$}

\date{} 
\begin{document}
\twocolumn[
\maketitle 
\vspace{-9mm}
\begin{center}
\begin{minipage}{1\textwidth}
\begin{center}
\textit{\textrm{
\textsuperscript{1} CAS Key Laboratory of Quantum Information, University of Science and Technology of China, Hefei 230026, China.
\\\textsuperscript{2} CAS Center For Excellence in Quantum Information and Quantum Physics, University of Science and Technology of China, Hefei, 230026, China.
\\\textsuperscript{3} Hefei National Laboratory, University of Science and Technology of China, Hefei 230088, China.
\\\textsuperscript{4} Department of Applied Physics, University of Geneva, 1211 Geneva, Switzerland.
\\\textsuperscript{5} Constructor Institute of Technology (CIT), Geneva, Switzerland.
\\\textsuperscript{6} Physics Department and NanoLund, Lund University, Box 118, 22100 Lund, Sweden.
\\\textsuperscript{7} These authors contributed equally: Chao Zhang, Jia-Le Miao.
\\\ ~~~Emails to: huxm@ustc.edu.cn; jef.pauwels@unige.ch; armin.tavakoli@teorfys.lu.se; bhliu@ustc.edu.cn}}
\end{center}
\end{minipage}
\end{center}

\setlength\parindent{12pt}
\begin{quotation}
\noindent 
{\bf{Entanglement has the ability to enhance the transmission of classical information over a quantum channel. However, fully harvesting this advantage typically requires complex entangling measurements, which are challenging to implement and scale with the system's size. In this work, we consider a natural quantum information primitive in which the message to be communicated is selected stochastically. We introduce a protocol that leverages high-dimensional entanglement to perform this task perfectly, without requiring quantum interference between particles at the measurement station. We experimentally demonstrate the protocol’s scalability in an optical setup using 8-dimensional entanglement and multi-outcome detection, providing a practical solution for stochastic communication and a robust method for certifying the dimensionality of entanglement in communication experiments. 
}} 
\end{quotation}]

\begin{figure}[htb]
\includegraphics[width=0.95\columnwidth]{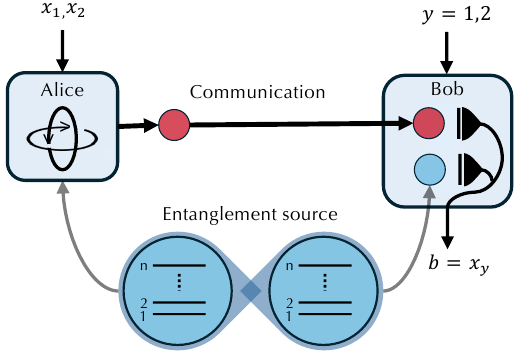}
\caption{\emph{Stochastic communication with a high-dimensional entanglement source.} Alice and Bob share an $n$-dimensional entangled state. Alice encodes her two possible messages $x = (x_1, x_2)$ by applying a unitary operation on her entangled particle, which is then sent to Bob. Bob indicates with $y$ the message of his interest and decodes it by performing separate measurements on his two particles.}\label{scenario}
\end{figure}

\noindent
The use of quantum resources for communication is one of the central promises of quantum information science and technology. Quantum advantages over classical communication can be achieved in two ways: by senders and receivers replacing classical messages with quantum ones \cite{Spekkens2009, Hendrych2012, Ahrens2012}, or by sharing an entangled state before employing classical communication \cite{Buhrman2010, Pawłowski2009, Muhammad2014}. In the first case, the advantage stems from quantum superposition \cite{Wiesner1983}, whilein the second, it is powered by Bell nonlocality \cite{Brukner2004, Pauwels2022}. However, the strongest quantum communication resource combines shared entanglement between sender and receiver with the transmission of quantum messages over the channel \cite{PhysRevLett.69.2881}. This is illustrated in Fig.~\ref{scenario}. Numerous experiments \cite{Mattle1996, Schuck2006, Barreiro2008, Williams2017} have demonstrated violations of classical limits using qubit entanglement and qubit messages. In theory, however,  quantum advantages become even more significant when the employed quantum resources are of higher-than-qubit dimension.

Accessing these advantages on photonic platforms, especially for high-dimensional systems, remains challenging. Although high-quality, high-dimensional entanglement has become increasingly available in recent years \cite{Erhard2020, Friis2019}, realizing its full potential typically requires implementing high-dimensional single-photon gates for message encoding and entangled measurements with multi-outcome detection for the receiver's decoding. These measurements, acting jointly on the receiver's two particles (see Fig.~\ref{scenario}), often rely on ancillary photons when implemented with linear optics \cite{Lutkenhaus1999, Loock2004}. Significant effort has been dedicated to achieving entangled measurements on two photonic qubits \cite{Michler1996, Kim2001, Schuck2006,Hu2018,Bayerbach2023}, but they have yet to be demonstrated beyond dimension three \cite{Luo2019, Hu2020}. Recently, however, it was shown that some communication tasks can be optimally performed using only single-particle measurements on the receiver's side \cite{Piveteau2022, Bakhshinezhad2024}, bypassing this  technological barrier. However, the known scenarios where this is feasible do not correspond to operationally relevant tasks, they are not always able to achieve a perfect performance even under ideal conditions and it is also not known how to scale the advantages beyond qubit systems.

Here, we show that quantum messages assisted by shared entanglement can be used to perform a natural communication task using only single-photon measurements perfectly. In this task, the sender possesses two pieces of information, and with limited communication, the receiver can retrieve whichever piece they are interested in, constituting stochastic message recovery. We show that the magnitude of quantum advantage scales rapidly with the Hilbert space dimension. This can also serve as a noise-robust certification of the genuine dimension of entanglement, known as the Schmidt number, without requiring knowledge of dimension-restricted quantum operations. To demonstrate the feasibility and scalability of our approach, we use photonic entanglement and quantum messages of dimension eight to implement the communication protocol. This entails the realization of programmable eight-dimensional unitary transformations for message encoding and genuine eight-outcome photonic measurements for decoding. We utilize the planar extension of the path degree of freedom (DOF), combined with polarization DOF for encoding high-dimensional entangled states \cite{Hu2020d32}. This greatly facilitates the implementation of high-dimensional unitary operations and measurements \cite{Hu2020} in which all eight outcomes are resolved in every round. We report a near-optimal communication success rate of 0.9729, surpassing the theoretical limit of any protocol that does not utilize entanglement by 44 percentage points. Furthermore, this success rate exceeds the performance of protocols based on seven-dimensional entanglement, providing a certificate of genuine eight-dimensional entanglement without the need to characterize any part of the experiment other than the channel degrees of freedom.

\section*{Stochastic communication}
Consider that Alice has a dataset which consists of two pieces of information, $x=(x_1,x_2)$, where $x_1,x_2\in\{0,\ldots,n-1\}\equiv [n]$ for some natural number $n$. Alice cannot send her full dataset to Bob because she is limited to encoding $x$ into a smaller message, $m$, which carries at most $\log_2(n)$ bits of information. In classical communication, this corresponds to $m$ having $n$ possible values ($m\in[n]$). Bob can privately select which of Alice's two elements he wants to recover, and he marks his choice with $y\in\{1,2\}$. By decoding Alice's message, he aims to output $b=x_y$; see Fig.~\ref{scenario}. The success rate of this task is given by 
\begin{equation}\label{successrate}
	\mathcal{S} = \frac{1}{2n^2} \sum_{x_1,x_2=0}^{n-1} \sum_{y=1}^2 p(b=x_y|x,y)\,,
\end{equation}
where $p(b=x_y|x,y)$ is the probability of Bob's guess $b$ corresponding to the correct value $x_y$. The goal of Alice and Bob is to achieve the highest  success rate possible. However, it is known that with classical communication, the success rate is fundamentally limited by $\mathcal{S}\leq \frac{1}{2}+\frac{1}{2n}$ \cite{Ambainis2024}. Thus, perfect stochastic communication is not only classically impossible but the success rate also decreases monotonically in the message size $n$.

It is natural to ask whether quantum resources can provide an advantage. Previous works have investigated the scenario both when Alice and Bob are allowed to share an entangled state to enhance their classical communication \cite{Pawlowski2010, Tavakoli2016, Chailloux2016, Pauwels2021} and when they substitute the classical $n$-valued message for a quantum $n$-dimensional message \cite{PhysRevLett.114.170502, Farkas2024, Ambainis2009, deVicente2019}. While both of these quantum approaches improve on the best classical protocol, the success rates still decrease monotonically with $n$. For example, in the latter case the best success rate is given by $\mathcal{S}=\frac{1}{2}+\frac{1}{2\sqrt{n}}$ \cite{PhysRevLett.114.170502}.

\subsection*{Optimal quantum protocol}
Quantum protocols can perform much better by combining shared entanglement with an $n$-dimensional quantum message. By employing the seminal dense coding protocol \cite{PhysRevLett.69.2881}, in its high-dimensional form, Alice and Bob can perfectly perform stochastic communication, i.e.~they achieve $\mathcal{S}=1$. The protocol requires Alice and Bob to share an $n$-dimensional maximally entangled state $\ket{\phi_n} =\frac{1}{\sqrt{n}}\sum_{i=0}^{n-1} \ket{ii}$. Then, Alice encodes $x$ via the unitary operation  $U_{x_1x_2} =  Z^{x_2}X^{x_1}$, where $X=\sum_{k=0}^{n-1}\ketbra{k+1}{k}$  and $Z=\sum_{k=0}^{n-1} e^{\frac{2\pi i }{n}k}\ketbra{k}{k}$ are the shift and clock operators, respectively. The joint state of the two particles obtained by Bob is then 
\begin{equation}
	\ket{\psi_{x_1x_2}} = Z^{x_2}X^{x_1} \otimes \mathbb{1} \ket{\phi_n} \, \label{eq:encoding},
\end{equation}
corresponding to the $n$-dimensional Bell basis. Thus, if Bob measures this basis, he can learn $x$ and afterward select $x_y$ as his output. 

However, the crucial drawback here is that implementing the $n$-dimensional Bell basis measurement on separate photons is exceptionally challenging beyond the lowest value of $n$. Interestingly, there exists a much simpler quantum strategy that still achieves perfect stochastic communication.  In this strategy, Alice and Bob again share $\ket{\phi_n}$ and Alice uses the same unitaries as in dense coding, but Bob substitutes his measurement so that it now has an explicit dependence on $y$. The key insight is that while the Bell basis in dense coding allows the receiver to simultaneously decode both elements $x_1$ and $x_2$, Bob only needs to access one of these elements per round in the stochastic task. Alice's encoding has encoded $x_1$ in the parity of the states $\{\ket{\psi_{x_1x_2}}\}$ and $x_2$ in their phase. Bob can access each of these properties by means of separate measurements on the two systems, corresponding to $Z\otimes Z$ and $X\otimes X$ for $y=1,2$, respectively. The outcome of each single system measurement can then be classically wired to make the final output of Bob. The procedure can be effectively described by the measurement operators $M_{b|1}=\sum_{a=0}^{n-1}\ketbra{t_{ab}}{t_{ab}}$ and $M_{b|2}=(F\otimes F^\dagger) M_{b|1}(F^\dagger \otimes F)$, where $\ket{t_{ab}}=\ket{a+b,a}$ and $F$ is the Fourier matrix defined by $F=\sum_{k,l=0}^{n-1} e^{\frac{2\pi i }{n}kl}\ketbra{k}{l}$. In Appendix \textcolor{red}{1} we prove that these unentangled measurements achieve a unit success rate in the task.

\begin{figure*}[htb]
\includegraphics [width= 1\textwidth]{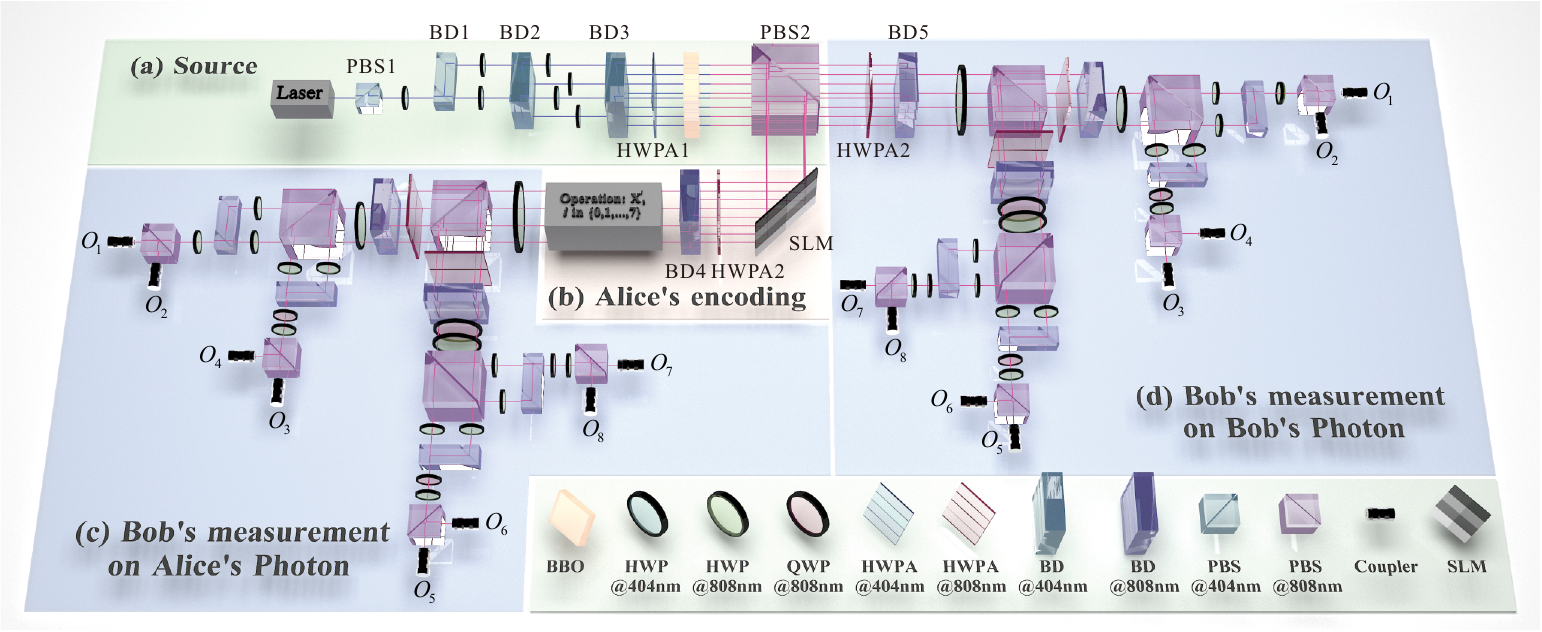}
\vspace{-0.3cm}%
\caption{\emph{Experimental setup.} (a): A continuous-wave (cw) laser operating at a wavelength 404~nm is split into 8 parallel paths by BD1-BD3, pumping the BBO crystal to undergo the type-II SPDC process to generate the correlated photon pairs (signal and idler photon) in 8-dimension Hilbert space. The photon pairs are separated by PBS2 and sent to Alice and Bob. (b): Alice perform unitary operation $U_{x_1x_2}$, including the clock unitary $Z^{x_2}$ and shift unitary $X^{x_1}$, on her photon and send it to Bob. (c) and (d): Note that the quantum state $|\Phi\rangle$ has been transformed to encode in path-polarization DOFs before Bob measures the photons. After Bob receives the photon from Alice, he performs the separate measurements $X\otimes X$ and $Z\otimes Z$ with multi-outcome based on his input $y=\{1,2\}$. Then, the coincidence events between two photons are determined by a time-to-digital converter (UQDevice). SLM: Spatial light modulator. HWP: Half wave plate. QWP: Quarter wave plate. HWPA: Half wave plate array. BD: Beam displacer. PBS: Polarizing beam splitter. BBO: $\beta$-barium-borate crystal.}
\label{setup}
\vspace{0.2cm}
\end{figure*}

\subsection*{Certifying high-dimensional entanglement}
The presented protocol relies on $n$-dimensional entanglement between Alice and Bob but it is natural to ask if this is truly necessary. It turns out that the entanglement dimension, formally known as the Schmidt number \cite{Terhal2000}, implies a simple limitation on the success rate of stochastic communication independently of the choices of quantum encoding and decoding operations. 

We are interested in the largest possible value of $\mathcal{S}$ achievable with $d$-dimensional entanglement. Since $\mathcal{S}$ is a linear expression we can w.l.g restrict ourselves to considering pure entangled states. For pure states, the Schmidt number reduces to the rank of the single-particle reductions of the state. Still, it is a hard problem to optimise $\mathcal{S}$ over all quantum encoding channels for Alice, all quantum measurements for Bob and all pure states with rank-$d$ marginals. In Appendix \textcolor{red}{2} we show how to derive an analytical upper bound on this quantity, namely
\begin{equation}\label{3}
	\mathcal{S}\leq \frac{1}{2}\qty(1+\sqrt{\frac{d}{n}} ) \,.
\end{equation}
Notice that for $d=1$ (corresponding to having no entanglement) it recovers the known optimal success rate for entanglement-unassisted quantum protocols and for $d=n$ it recovers the unit success rate saturated by our protocol presented above. For $1<d<n$, the bound is not expected to be optimal but numerical case studies indicate that it is often quite accurate.

Eq.~\eqref{3} shows that larger Schmidt numbers make possible larger success rates. This can also be interpreted as an entanglement certification tool: experiments that observe a sufficiently high value of $\mathcal{S}$ can conclude a lower bound on the Schmidt number, solely under the assumption of the system operating on $n$ degrees of freedom. Note that this requires no further prior characterization of the state, the encoding channels or the measurements.

\section*{Experimental realization}
\begin{figure}[htb]
\includegraphics [width= 1\columnwidth]{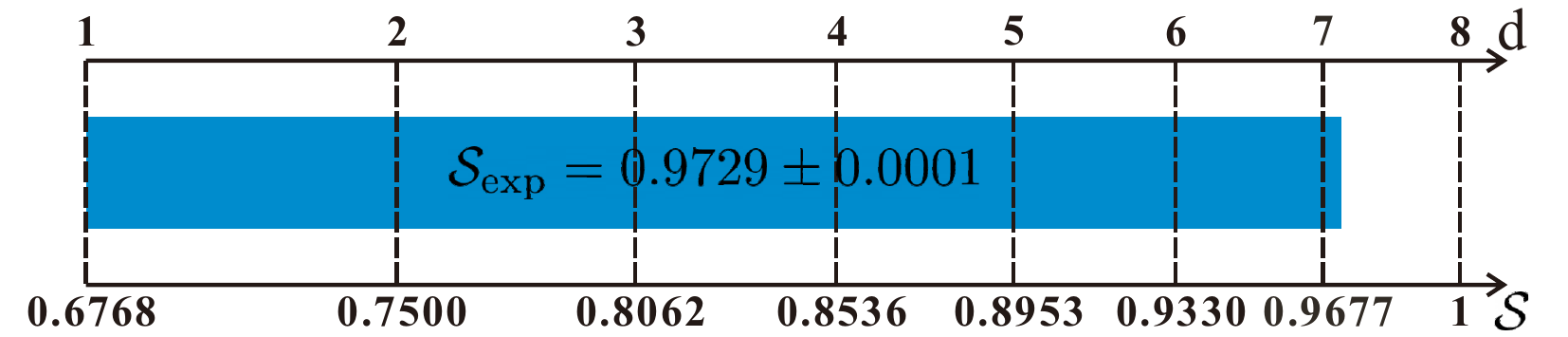}
\vspace{-0.3cm}%
\caption{Upper bounds on the stochastic communication rate, $\mathcal{S}$, for a quantum channel supporting a system of dimension $n=8$ and assisted by entanglement with Schmidt number $d$. We experimentally verify a Schmidt number of 8 by achieving a success rate of $\mathcal{S}_{\text{exp}} = 0.9729$ with an error of $\sigma = 1 \times 10^{-4}$, which surpasses the maximum achievable success rate of 0.9677 for a Schmidt number of 7.}
\label{bounds}
\vspace{0.2cm}
\end{figure}

We demonstrate the stochastic communication rate, the scalability of our protocol to high-dimensional messages, and its ability to detect entanglement. Our demonstration is based on an eight-dimensional quantum channel ($n=8$) assisted by an eight-dimensional entangled state. Following our theoretical discussion, an ideal implementation of the protocol using high-dimensional entanglement achieves a unit success rate. In practice, this rate is reduced because of noise and imperfections. In Fig.~\ref{bounds}, we list the theoretical limitation on the success rate implied by Eq.~\eqref{3} for any possible protocol based on the entanglement dimension $d\leq n$.

\subsection*{Experimental scheme}

The experiment involves three main parts; see Fig.~\ref{setup} for an illustration. These parts are (i) the generation of two-photon  entangled states of high dimensionality, (ii) the application of high-dimensional unitaries on one photon in order to encode the messages, and (iii) a multi-outcome detection scheme implemented individually on the two photons at the measuring station. 
In the following, we discuss each of these parts separately.

{\textbf{(i) Generation of 8-dimensional entangled states.}} We prepared 8-dimensional entangled states encoded in the polarization and path DOFs in-plane. 
A 404~CW laser is polarized by PBS1 (cf. Fig.~\ref{setup}a) and then separated into eight paths with horizontal polarization, forming a $4\times2$ array after passing through BD1-BD3.  Specifically, BD1 (BD2) refracts a vertically (horizontally) polarized laser by 4~mm in the horizontal (vertical) direction, and BD3 refracts horizontally polarized laser by 2~mm in the vertical direction. These eight equal beams pump a 4-mm-thick BBO crystal, undergoing a type-II spontaneous parametric down-conversion (SPDC) process~\cite{Schneeloch_2019}, resulting in two correlated photons 
\begin{equation}
   \left|\Phi\right\rangle=\frac{1}{\sqrt{8}} \sum_{k=0}^{7}|k_H\rangle |k_V\rangle, 
\end{equation}
with different polarization in 8-dimensional Hilbert space. The photons with horizontal polarization are distributed to Alice and the others to Bob, resulting in a shared entangled state.

{\textbf{(ii) Encoding, conversion and high-dimensional local operations.}} The initial encoding of the entangled states is in path DOF with planar extension and Alice utilizes the SLM segmented into 8 regions to independently modulate the phase of each subspace, thereby implementing the high-dimensional clock operation \(Z^{x_2}\)~\cite{PhysRevLett.125.090503}. HWPA2 was then used to adjust the polarization of each path, while BD4 reflected the paths in rows 1 and 3 downward by 2 mm, forming a \(2 \times 2\) matrix with two polarization states in each path. Similarly, HWPA3 and BD5 performed the same operations for Bob's decoding. At this point, the encoding of the entangled states is transformed from path DOF to path-polarization DOFs and the shift operation \(X^{x_1}\) is constructed using a series of BDs and HWPAs (cf. Fig.~\ref{setup}b). Note that each optical element can operate multiple subspaces simultaneously, demonstrating that our encoding method efficiently performs high-dimensional operations compared to the encoding in path-polarization DOFs within a single dimension. Details are provided in Supplementary Materials. Essentially, after Alice's local operation, we generate 64 Bell states within an 8-dimensional Hilbert space. Alice then transmits her quantum states to Bob through a quantum communication channel with \(n = 8\).

{\textbf{(iii) Multi-outcome high-dimensional measurement.}} Bob selects the appropriate measurement, $M_{b|1}$ or $M_{b|2}$, based on his input $y \in {1, 2}$. The first measurement ($y = 1$) corresponds to a post-processing procedure, where each particle is separately measured in the computational basis. Similarly, the second measurement ($y = 2$) involves post-processing based on independent measurements of each particle in the Fourier basis. Thus, the two measurements correspond to two pairs of mutually unbiased bases (MUBs), namely $Z \otimes Z$ and $X \otimes X$. Using the path-polarization degrees of freedom (DOFs) for encoding, we construct the multi-outcome operators $O_1$ through $O_8$ as the eigenvectors of the high-dimensional measurements $X$ and $Z$, respectively (detailed in Appendix \textcolor{red}{5}, cf. Fig.\ref{setup}cd). Finally, the measured photons pass through bandpass filters with a bandwidth of 3 nm, are collected in couplers, and are detected by an avalanche photodiode (APD). From many repetitions of the protocol using different settings, we estimate the success rate $\mathcal{S}_{exp}$, from Eq.~(\ref{successrate}).

\subsection*{Results}

In our experiment, the brightness of the prepared shared entangled state \(|\Phi\rangle\) was approximately 800 Hz, with an integration time of 100 s for each measurement setting $\{M_{b|1}\}$ or $\{M_{b|2}\}$. Each two-dimensional subspace of the prepared entangled state, such as \(\{|0\rangle, |1\rangle\}\), was measured using the computational basis \(\{|0\rangle, |1\rangle\}\) and the Fourier basis \(\{(|0\rangle+|1\rangle)/\sqrt{2}, (|0\rangle-|1\rangle)/\sqrt{2}\}\), achieving visibilities of \(0.999 \pm 0.001\) and \(0.990 \pm 0.001\) respectively.

Based on our experimental scheme, we ran all combinations of Alice's encodings $(x_1,x_2)$, and performed product measurements $\{M_{b|1}\}$ and $\{M_{b|2}\}$ and recorded the successful events $x_y=b$. From this, we estimate the probabilities $p(b=x_y|x_1,x_2,y)$ and a total success rate of
\begin{equation}  
\mathcal{S}_{\text{exp}} = 0.9729 \pm 0.0001 \,, 
\label{result}  
\end{equation}
 where the statistical error of $10^{-4}$ was determined by simulating the measured photons using 1000 samples from a Poisson distribution~\cite{harrison2010introduction,joy1991introduction}. Details of the grayscale value calibration and experimental statistical analysis are provided in Appendix \textcolor{red}{6}. Our experimental result exceeds the bound \(\mathcal{S} = 0.9677\) for any protocol based on seven-dimensional entanglement, see Fig.~\ref{bounds}. This allows us to certify that Alice and Bob share an entangled state with Schmidt number $d=8$, with vanishingly small $p$-value, see Appendix \textcolor{red}{7}.

\section*{Discussion}

The seminal dense coding protocol \cite{PhysRevLett.69.2881} demonstrated that entanglement enhances communication over a quantum channel. However, its implementation relies on performing a complete Bell state measurement on a pair of photons, each encoding a $d$-dimensional quantum system. Scaling such measurements for large messages is highly challenging, as they would require numerous ancillary photons in linear optics setups. In this work, we have shown that these limitations can be mitigated when the message to be communicated is selected stochastically. This form of stochastic messaging serves as a fundamental building block for various quantum information applications, such as randomness generation \cite{PhysRevA.85.052308}, quantum key distribution \cite{Pawlowski2011, Woodhead2015}, quantum certification \cite{PhysRevA.98.062307,Tavakoli2020c}, and first-principles derivations of quantum theory predictions \cite{Pawlowski2009}. Our results show that stochastic communication can be performed deterministically, without the need for entangling operations or even photon-photon interference at the measurement station. This opens the door to scalable experimental implementations, which we have demonstrated using a pair of 8-dimensional entangled photons for stochastically transmitting a pair of 8-symbol messages.

To address the challenges associated with generating and manipulating high-dimensional systems, our experiment utilizes the planar extension of the path degree of freedom (DOF) combined with the polarization DOF to encode high-dimensional entangled states. This method allows for the simultaneous manipulation of multiple subspaces using a single optical component, greatly simplifying the implementation of high-dimensional local operations and multi-outcome measurements. 
Notably, this encoding method offers effective scalability~\cite{PhysRevLett.125.090503} and can be integrated with multi-core fibers for long-distance distribution~\cite{Hu:20,ding2017high,PhysRevA.96.022317}, establishing a solid foundation for practical field applications. Furthermore, while many experiments on high-dimensional quantum correlations rely on post-processing data from multiple separate projections, our experiment resolves all possible outcomes in each round. Such multi-outcome detection schemes are crucial for closing loopholes that could otherwise undermine the validity of the conclusions drawn from these experiments \cite{Tavakoli2024b}.

In addition, we have shown that the transmission rate of stochastic communication  can serve as a detector of the dimensionality of the shared entangled state. Importantly, the dimensionality is inferred without making assumptions about the inner workings of any part of the experiment; the only knowledge needed is the specific DOF controlled for the system transmitted over the communication line. By achieving a near-optimal transmission rate, we were able to certify the creation of a genuinely 8-dimensional entangled state. Alternative black-box-type methods for such certification are based on Bell nonlocality and quantum steering. For the former, noise requirements are typically significantly higher and no general formulas are known for certifying arbitrary-dimensional entanglement.  For the latter, stringent assumptions are necessary, as one must have perfect control over one of the measurement devices \cite{Designolle2021, Gois2024, dAlessandro2024}. Even small deviations from this idealization can result in false positives \cite{Morelli2022, Sarkar2023, Tavakoli2024}.

\printbibliography

@article{Azuma1967,
author = {Kazuoki Azuma},
title = {{Weighted sums of certain dependent random variables}},
volume = {19},
journal = {Tohoku Mathematical Journal},
number = {3},
publisher = {Tohoku University, Mathematical Institute},
pages = {357 -- 367},
year = {1967},
doi = {10.2748/tmj/1178243286},
URL = {https://doi.org/10.2748/tmj/1178243286}
}

@Article{Gill2002,
  author  = {Gill, Richard D.},
  journal = {Foundations of Probability and Physics},
  title   = {Time, Finite Statistics, and Bell's Fifth Position},
  year    = {2002},
  pages   = {179--206},
  volume  = {5},
  doi     = {10.1.1.6.5788},
    url = {https://doi.org/10.1.1.6.5788}
}

@ARTICLE{Piveteau2022,
       author = {{Piveteau}, Am{\'e}lie and {Pauwels}, Jef and {H{\r a}kansson}, Emil and {Muhammad}, Sadiq and {Bourennane}, Mohamed and {Tavakoli}, Armin},
        title = "{Entanglement-assisted quantum communication with simple measurements}",
      journal = {Nature Communications},
         year = 2022,
        month = {12},
       volume = {13},
          eid = {7878},
        pages = {7878},
          doi = {10.1038/s41467-022-33922-5},
       url = {https://doi.org/10.1038/s41467-022-33922-5}
}

@Article{Pauwels2022,
  author    = {Pauwels, Jef and Pironio, Stefano and Cruzeiro, Emmanuel Zambrini and Tavakoli, Armin},
  journal   = {Phys. Rev. Lett.},
  title     = {Adaptive Advantage in Entanglement-Assisted Communications},
  year      = {2022},
  month     = {9},
  pages     = {120504},
  volume    = {129},
  doi       = {10.1103/PhysRevLett.129.120504},
  issue     = {12},
  numpages  = {7},
  publisher = {American Physical Society},
  url       = {https://doi.org/10.1103/PhysRevLett.129.120504},
}

@article{PhysRevA.99.032316,
  title = {Self-testing mutually unbiased bases in the prepare-and-measure scenario},
  author = {Farkas, M\'at\'e and Kaniewski, J\ifmmode \mbox{\k{e}}\else \k{e}\fi{}drzej},
  journal = {Phys. Rev. A},
  volume = {99},
  issue = {3},
  pages = {032316},
  numpages = {11},
  year = {2019},
  month = {Mar},
  publisher = {American Physical Society},
  doi = {10.1103/PhysRevA.99.032316},
  url = {https://doi.org/10.1103/PhysRevA.99.032316}
}

@Article{Pawłowski2009,
	author={Paw{\l}owski, Marcin
	and Paterek, Tomasz
	and Kaszlikowski, Dagomir
	and Scarani, Valerio
	and Winter, Andreas
	and {\.{Z}}ukowski, Marek},
	title={Information causality as a physical principle},
	journal={Nature},
	year={2009},
	month={10},
	day={01},
	volume={461},
	number={7267},
	pages={1101-1104},
	issn={1476-4687},
	doi={10.1038/nature08400},
	url={https://doi.org/10.1038/nature08400}
}

@article{SDPreview,
  title = {Semidefinite programming relaxations for quantum correlations},
  author = {Tavakoli, Armin and Pozas-Kerstjens, Alejandro and Brown, Peter and Ara\'ujo, Mateus},
  journal = {Rev. Mod. Phys.},
  volume = {96},
  issue = {4},
  pages = {045006},
  numpages = {68},
  year = {2024},
  month = {12},
  publisher = {American Physical Society},
  doi = {10.1103/RevModPhys.96.045006},
  url = {https://doi.org/10.1103/RevModPhys.96.045006}
}

@article{Muhammad2014,
	title = {Quantum Bidding in Bridge},
	author = {Muhammad, Sadiq and Tavakoli, Armin and Kurant, Maciej and Paw\l{}owski, Marcin and \ifmmode \dot{Z}\else \.{Z}\fi{}ukowski, Marek and Bourennane, Mohamed},
	journal = {Phys. Rev. X},
	volume = {4},
	issue = {2},
	pages = {021047},
	numpages = {10},
	year = {2014},
	month = {6},
	publisher = {American Physical Society},
	doi = {10.1103/PhysRevX.4.021047},
	url = {https://doi.org/10.1103/PhysRevX.4.021047}
}

@article{PhysRevLett.114.170502,
  title = {Quantum Random Access Codes Using Single $d$-Level Systems},
  author = {Tavakoli, Armin and Hameedi, Alley and Marques, Breno and Bourennane, Mohamed},
  journal = {Phys. Rev. Lett.},
  volume = {114},
  issue = {17},
  pages = {170502},
  numpages = {5},
  year = {2015},
  month = {4},
  publisher = {American Physical Society},
  doi = {10.1103/PhysRevLett.114.170502},
  url = {https://doi.org/10.1103/PhysRevLett.114.170502}
}

@Article{Pauwels2021,
       author = {{Pauwels}, Jef and {Tavakoli}, Armin and {Woodhead}, Erik and {Pironio}, Stefano},
        title = "{Entanglement in prepare-and-measure scenarios: many questions, a few answers}",
      journal = {New Journal of Physics},
         year = 2022,
        month = {6},
       volume = {24},
       number = {6},
          eid = {063015},
        pages = {063015},
          doi = {10.1088/1367-2630/ac724a},
  url = {https://doi.org/10.1088/1367-2630/ac724a}
}

@article{Woodhead2015,
  title = {Secrecy in Prepare-and-Measure Clauser-Horne-Shimony-Holt Tests with a Qubit Bound},
  author = {Woodhead, Erik and Pironio, Stefano},
  journal = {Phys. Rev. Lett.},
  volume = {115},
  issue = {15},
  pages = {150501},
  numpages = {5},
  year = {2015},
  month = {10},
  publisher = {American Physical Society},
  doi = {10.1103/PhysRevLett.115.150501},
  url = {https://doi.org/10.1103/PhysRevLett.115.150501}
}

@Article{Pauwels2022b,
  author        = {Pauwels, Jef and Pironio, Stefano and Woodhead, Erik and Tavakoli, Armin},
  journal       = {Phys. Rev. Lett.},
  title         = {Almost Qudits in the Prepare-and-Measure Scenario},
  year          = {2022},
  month         = {12},
  pages         = {250504},
  volume        = {129},
  doi           = {10.1103/PhysRevLett.129.250504},
  issue         = {25},
  numpages      = {7},
  publisher     = {American Physical Society},
  url           = {https://doi.org/10.1103/PhysRevLett.129.250504},
}

@article{PhysRevLett.69.2881,
  title = {Communication via one- and two-particle operators on Einstein-Podolsky-Rosen states},
  author = {Bennett, Charles H. and Wiesner, Stephen J.},
  journal = {Phys. Rev. Lett.},
  volume = {69},
  issue = {20},
  pages = {2881--2884},
  numpages = {0},
  year = {1992},
  month = {11},
  publisher = {American Physical Society},
  doi = {10.1103/PhysRevLett.69.2881},
  url = {https://doi.org/10.1103/PhysRevLett.69.2881}
}

@Article{Tavakoli2016,
  author    = {Tavakoli, Armin and Marques, Breno and Paw\l{}owski, Marcin and Bourennane, Mohamed},
  journal   = {Phys. Rev. A},
  title     = {Spatial versus sequential correlations for random access coding},
  year      = {2016},
  month     = {3},
  pages     = {032336},
  volume    = {93},
  doi       = {10.1103/PhysRevA.93.032336},
  issue     = {3},
  numpages  = {6},
  publisher = {American Physical Society},
  url       = {https://doi.org/10.1103/PhysRevA.93.032336},
}

@article{Morelli2022,
  title = {Entanglement Detection with Imprecise Measurements},
  author = {Morelli, Simon and Yamasaki, Hayata and Huber, Marcus and Tavakoli, Armin},
  journal = {Phys. Rev. Lett.},
  volume = {128},
  issue = {25},
  pages = {250501},
  numpages = {6},
  year = {2022},
  month = {6},
  publisher = {American Physical Society},
  doi = {10.1103/PhysRevLett.128.250501},
  url = {https://doi.org/10.1103/PhysRevLett.128.250501}
}

@Article{Kittaneh1997,
  author    = {Kittaneh, Fuad},
  journal   = {Journal of Functional Analysis},
  title     = {Norm Inequalities for Certain Operator Sums},
  year      = {1997},
  issn      = {0022-1236},
  month     = feb,
  number    = {2},
  pages     = {337--348},
  volume    = {143},
  doi       = {10.1006/jfan.1996.2957},
  publisher = {Elsevier BV},
url = {https://doi.org/doi/10.1006/jfan.1996.2957}
}

@article{Gois2024,
  title = {Complete Hierarchy for High-Dimensional Steering Certification},
  author = {de Gois, Carlos and Pl\'avala, Martin and Schwonnek, Ren\'e and G\"uhne, Otfried},
  journal = {Phys. Rev. Lett.},
  volume = {131},
  issue = {1},
  pages = {010201},
  numpages = {6},
  year = {2023},
  month = {7},
  publisher = {American Physical Society},
  doi = {10.1103/PhysRevLett.131.010201},
  url = {https://doi.org/10.1103/PhysRevLett.131.010201}
}

@misc{dAlessandro2024,
      title={Semidefinite relaxations for high-dimensional entanglement in the steering scenario}, 
      author={Nicola D'Alessandro and Carles Roch i Carceller and Armin Tavakoli},
      year={2024},
      eprint={2410.02554},
      archivePrefix={arXiv},
      primaryClass={quant-ph},
      url={https://arxiv.org/abs/2410.02554}, 
}

@Unpublished{Tavakoli2024b,
      title={The binarisation loophole in high-dimensional quantum correlation experiments}, 
      author={Armin Tavakoli and Roope Uola and Jef Pauwels},
      year={2024},
      eprint={2407.16305},
      archivePrefix={arXiv},
      primaryClass={quant-ph},
      url={https://arxiv.org/abs/2407.16305}, 
}

@Article{Pawlowski2009,
  author    = {Paw{\l}owski, Marcin and Paterek, Tomasz and Kaszlikowski, Dagomir and Scarani, Valerio and Winter, Andreas and {\.Z}ukowski, Marek},
  journal   = {Nature},
  title     = {Information causality as a physical principle},
  year      = {2009},
  issn      = {1476-4687},
  month     = {10},
  number    = {7267},
  pages     = {1101--1104},
  volume    = {461},
  day       = {01},
  doi       = {10.1038/nature08400},
  publisher = {Nature Publishing Group},
url = {https://doi.org/10.1038/nature08400}
}

@Article{Tavakoli2020c,
  author    = {Tavakoli, Armin and Smania, Massimiliano and V{\'e}rtesi, Tam{\'a}s and Brunner, Nicolas and Bourennane, Mohamed},
  journal   = {Sci.\ Adv.},
  title     = {Self-testing nonprojective quantum measurements in prepare-and-measure experiments},
  year      = {2020},
  number    = {16},
  pages     = {eaaw6664},
  volume    = {6},
  doi       = {10.1126/sciadv.aaw6664},
  publisher = {American Association for the Advancement of Science},
  url = {https://doi.org/10.1126/sciadv.aaw6664}
}

@article{PhysRevA.85.052308,
  title = {Semi-device-independent randomness certification using $n\ensuremath{\rightarrow}1$ quantum random access codes},
  author = {Li, Hong-Wei and Paw\l{}owski, Marcin and Yin, Zhen-Qiang and Guo, Guang-Can and Han, Zheng-Fu},
  journal = {Phys. Rev. A},
  volume = {85},
  issue = {5},
  pages = {052308},
  numpages = {4},
  year = {2012},
  month = {5},
  publisher = {American Physical Society},
  doi = {10.1103/PhysRevA.85.052308},
  url = {https://doi.org/10.1103/PhysRevA.85.052308}
}

@Article{Terhal2000,
  author    = {Terhal, Barbara M. and Horodecki, Paweł},
  journal   = {Physical Review A},
  title     = {Schmidt number for density matrices},
  year      = {2000},
  issn      = {1094-1622},
  month     = mar,
  number    = {4},
  pages     = {040301},
  volume    = {61},
  doi       = {10.1103/physreva.61.040301},
  publisher = {American Physical Society (APS)},
 url       = {https://doi.org/10.1103/physreva.61.040301},
}

@Article{Tavakoli2024,
  author    = {Tavakoli, Armin},
  journal   = {Physical Review Letters},
  title     = {Quantum Steering with Imprecise Measurements},
  year      = {2024},
  month     = feb,
  number    = {7},
  pages     = {070204},
  volume    = {132},
  doi       = {10.1103/physrevlett.132.070204},
  publisher = {American Physical Society (APS)},
  url = {https://doi.org/10.1103/physrevlett.132.070204}
}

@Article{Sarkar2023,
 title={Distrustful quantum steering},
  author={Sarkar, Shubhayan},
  journal={Physical Review A},
  volume={108},
  number={4},
  pages={L040401},
  year={2023},
  publisher={APS},
  doi = {10.1103/PhysRevA.108.L040401},
  url = {https://doi.org/10.1103/PhysRevA.108.L040401}
}

@article{Wiesner1983,
	author = {Wiesner, Stephen},
	title = {Conjugate coding},
	year = {1983},
	publisher = {Association for Computing Machinery},
	address = {New York, NY, USA},
	volume = {15},
	number = {1},
	issn = {0163-5700},
	doi = {10.1145/1008908.1008920},
	journal = {SIGACT News},
	month = {1},
	pages = {78–88},
	numpages = {11},
url = {https://doi.org/10.1145/1008908.1008920}
}

@article{Pawlowski2011,
  title = {Semi-device-independent security of one-way quantum key distribution},
  author = {Paw\l{}owski, Marcin and Brunner, Nicolas},
  journal = {Phys. Rev. A},
  volume = {84},
  issue = {1},
  pages = {010302},
  numpages = {4},
  year = {2011},
  month = {7},
  publisher = {American Physical Society},
  doi = {10.1103/PhysRevA.84.010302},
  url = {https://doi.org/10.1103/PhysRevA.84.010302}
}

@article{PhysRevA.98.062307,
  title = {Self-testing quantum states and measurements in the prepare-and-measure scenario},
  author = {Tavakoli, Armin and Kaniewski, J\ifmmode \mbox{\k{e}}\else \k{e}\fi{}drzej and V\'ertesi, Tam\'as and Rosset, Denis and Brunner, Nicolas},
  journal = {Phys. Rev. A},
  volume = {98},
  issue = {6},
  pages = {062307},
  numpages = {13},
  year = {2018},
  month = {12},
  publisher = {American Physical Society},
  doi = {10.1103/PhysRevA.98.062307},
  url = {https://doi.org/10.1103/PhysRevA.98.062307}
}

@article{Spekkens2009,
	title = {Preparation Contextuality Powers Parity-Oblivious Multiplexing},
	author = {Spekkens, Robert W. and Buzacott, D. H. and Keehn, A. J. and Toner, Ben and Pryde, G. J.},
	journal = {Phys. Rev. Lett.},
	volume = {102},
	issue = {1},
	pages = {010401},
	numpages = {4},
	year = {2009},
	month = {1},
	publisher = {American Physical Society},
	doi = {10.1103/PhysRevLett.102.010401},
	url = {https://doi.org/10.1103/PhysRevLett.102.010401}
}

@Article{Hendrych2012,
	author={Hendrych, Martin
	and Gallego, Rodrigo
	and Mi{\v{c}}uda, Michal
	and Brunner, Nicolas
	and Ac{\'i}n, Antonio
	and Torres, Juan P.},
	title={Experimental estimation of the dimension of classical and quantum systems},
	journal={Nature Physics},
	year={2012},
	month={8},
	day={01},
	volume={8},
	number={8},
	pages={588-591},
	issn={1745-2481},
	doi={10.1038/nphys2334},
	url={https://doi.org/10.1038/nphys2334}
}

@Article{Ahrens2012,
	author={Ahrens, Johan
	and Badziag, Piotr
	and Cabello, Ad{\'a}n
	and Bourennane, Mohamed},
	title={Experimental device-independent tests of classical and quantum dimensions},
	journal={Nature Physics},
	year={2012},
	month={8},
	day={01},
	volume={8},
	number={8},
	pages={592-595},
	issn={1745-2481},
	doi={10.1038/nphys2333},
	url={https://doi.org/10.1038/nphys2333}
}

@article{Brukner2004,
	title = {Bell's Inequalities and Quantum Communication Complexity},
	author = {Brukner, \ifmmode \check{C}\else \v{C}\fi{}aslav and \ifmmode \dot{Z}\else \.{Z}\fi{}ukowski, Marek and Pan, Jian-Wei and Zeilinger, Anton},
	journal = {Phys. Rev. Lett.},
	volume = {92},
	issue = {12},
	pages = {127901},
	numpages = {4},
	year = {2004},
	month = {3},
	publisher = {American Physical Society},
	doi = {10.1103/PhysRevLett.92.127901},
	url = {https://doi.org/10.1103/PhysRevLett.92.127901}
}

@Article{Friis2019,
	author={Friis, Nicolai
	and Vitagliano, Giuseppe
	and Malik, Mehul
	and Huber, Marcus},
	title={Entanglement certification from theory to experiment},
	journal={Nature Reviews Physics},
	year={2019},
	month={1},
	day={01},
	volume={1},
	number={1},
	pages={72-87},
	issn={2522-5820},
	doi={10.1038/s42254-018-0003-5},
	url={https://doi.org/10.1038/s42254-018-0003-5}
}

@article{Lutkenhaus1999,
	title = {Bell measurements for teleportation},
	author = {L\"utkenhaus, N. and Calsamiglia, J. and Suominen, K.-A.},
	journal = {Phys. Rev. A},
	volume = {59},
	issue = {5},
	pages = {3295--3300},
	numpages = {0},
	year = {1999},
	month = {5},
	publisher = {American Physical Society},
	doi = {10.1103/PhysRevA.59.3295},
	url = {https://doi.org/10.1103/PhysRevA.59.3295}
}

@article{Loock2004,
	title = {Simple criteria for the implementation of projective measurements with linear optics},
	author = {van Loock, Peter and L\"utkenhaus, Norbert},
	journal = {Phys. Rev. A},
	volume = {69},
	issue = {1},
	pages = {012302},
	numpages = {5},
	year = {2004},
	month = {1},
	publisher = {American Physical Society},
	doi = {10.1103/PhysRevA.69.012302},
	url = {https://doi.org/10.1103/PhysRevA.69.012302}
}

@article{Hu2020,
	title = {Experimental High-Dimensional Quantum Teleportation},
	author = {Hu, Xiao-Min and Zhang, Chao and Liu, Bi-Heng and Cai, Yu and Ye, Xiang-Jun and Guo, Yu and Xing, Wen-Bo and Huang, Cen-Xiao and Huang, Yun-Feng and Li, Chuan-Feng and Guo, Guang-Can},
	journal = {Phys. Rev. Lett.},
	volume = {125},
	issue = {23},
	pages = {230501},
	numpages = {6},
	year = {2020},
	month = {12},
	publisher = {American Physical Society},
	doi = {10.1103/PhysRevLett.125.230501},
	url = {https://doi.org/10.1103/PhysRevLett.125.230501}
}

@article{Luo2019,
	title = {Quantum Teleportation in High Dimensions},
	author = {Luo, Yi-Han and Zhong, Han-Sen and Erhard, Manuel and Wang, Xi-Lin and Peng, Li-Chao and Krenn, Mario and Jiang, Xiao and Li, Li and Liu, Nai-Le and Lu, Chao-Yang and Zeilinger, Anton and Pan, Jian-Wei},
	journal = {Phys. Rev. Lett.},
	volume = {123},
	issue = {7},
	pages = {070505},
	numpages = {6},
	year = {2019},
	month = {8},
	publisher = {American Physical Society},
	doi = {10.1103/PhysRevLett.123.070505},
	url = {https://doi.org/10.1103/PhysRevLett.123.070505}
}

@article{Mattle1996,
	title = {Dense Coding in Experimental Quantum Communication},
	author = {Mattle, Klaus and Weinfurter, Harald and Kwiat, Paul G. and Zeilinger, Anton},
	journal = {Phys. Rev. Lett.},
	volume = {76},
	issue = {25},
	pages = {4656--4659},
	numpages = {0},
	year = {1996},
	month = {6},
	publisher = {American Physical Society},
	doi = {10.1103/PhysRevLett.76.4656},
	url = {https://doi.org/10.1103/PhysRevLett.76.4656}
}

@article{Pawlowski2010,
	title = {Entanglement-assisted random access codes},
	author = {Paw\l{}owski, Marcin and \ifmmode \dot{Z}\else \.{Z}\fi{}ukowski, Marek},
	journal = {Phys. Rev. A},
	volume = {81},
	issue = {4},
	pages = {042326},
	numpages = {4},
	year = {2010},
	month = {4},
	publisher = {American Physical Society},
	doi = {10.1103/PhysRevA.81.042326},
	url = {https://doi.org/10.1103/PhysRevA.81.042326}
}

@article{Chailloux2016,
	doi = {10.1088/1367-2630/18/4/045003},
	url = {https://dx.doi.org/10.1088/1367-2630/18/4/045003},
	year = {2016},
	month = {4},
	publisher = {IOP Publishing},
	volume = {18},
	number = {4},
	pages = {045003},
	author = {André Chailloux and Iordanis Kerenidis and Srijita Kundu and Jamie Sikora},
	title = {Optimal bounds for parity-oblivious random access codes},
	journal = {New Journal of Physics}
}

@misc{Ambainis2024,
	title={Quantum Advantages in (n,d)->1 Random Access Codes}, 
	author={Andris Ambainis and Dmitry Kravchenko and Sk Sazim and Joonwoo Bae and Ashutosh Rai},
	year={2024},
	eprint={1510.03045},
	archivePrefix={arXiv},
	primaryClass={quant-ph},
	url={https://arxiv.org/abs/1510.03045}, 
}

@Article{Buhrman2010,
  author    = {Buhrman, Harry and Cleve, Richard and Massar, Serge and de Wolf, Ronald},
  journal   = {Rev. Mod. Phys.},
  title     = {Nonlocality and communication complexity},
  year      = {2010},
  issn      = {1539-0756},
  month     = {3},
  number    = {1},
  pages     = {665--698},
  volume    = {82},
  doi       = {10.1103/RevModPhys.82.665},
  issue     = {1},
  numpages  = {0},
  publisher = {American Physical Society},
  url       = {https://doi.org/10.1103/RevModPhys.82.665},
}

@article{Designolle2021,
	title = {Genuine High-Dimensional Quantum Steering},
	author = {Designolle, S\'ebastien and Srivastav, Vatshal and Uola, Roope and Valencia, Natalia Herrera and McCutcheon, Will and Malik, Mehul and Brunner, Nicolas},
	journal = {Phys. Rev. Lett.},
	volume = {126},
	issue = {20},
	pages = {200404},
	numpages = {7},
	year = {2021},
	month = {5},
	publisher = {American Physical Society},
	doi = {10.1103/PhysRevLett.126.200404},
	url = {https://doi.org/10.1103/PhysRevLett.126.200404}
}

@article{deVicente2019,
	doi = {10.1088/1751-8121/aafde7},
	url = {https://dx.doi.org/10.1088/1751-8121/aafde7},
	year = {2019},
	month = {2},
	publisher = {IOP Publishing},
	volume = {52},
	number = {9},
	pages = {095304},
	author = {Julio I de Vicente},
	title = {A general bound for the dimension of quantum behaviours in the prepare-and-measure scenario},
	journal = {Journal of Physics A: Mathematical and Theoretical}
}

@misc{Farkas2024,
	title={Simple and general bounds on quantum random access codes}, 
	author={Máté Farkas and Nikolai Miklin and Armin Tavakoli},
	year={2024},
	eprint={2312.14142},
	archivePrefix={arXiv},
	primaryClass={quant-ph},
	url={https://arxiv.org/abs/2312.14142}, 
}

@misc{Ambainis2009,
	title={Quantum Random Access Codes with Shared Randomness}, 
	author={Andris Ambainis and Debbie Leung and Laura Mancinska and Maris Ozols},
	year={2009},
	eprint={0810.2937},
	archivePrefix={arXiv},
	primaryClass={quant-ph},
	url={https://arxiv.org/abs/0810.2937}, 
}

@article{Michler1996,
	title = {Interferometric Bell-state analysis},
	author = {Michler, Markus and Mattle, Klaus and Weinfurter, Harald and Zeilinger, Anton},
	journal = {Phys. Rev. A},
	volume = {53},
	issue = {3},
	pages = {R1209--R1212},
	numpages = {0},
	year = {1996},
	month = {3},
	publisher = {American Physical Society},
	doi = {10.1103/PhysRevA.53.R1209},
	url = {https://doi.org/10.1103/PhysRevA.53.R1209}
}

@article{Kim2001,
	title = {Quantum Teleportation of a Polarization State with a Complete Bell State Measurement},
	author = {Kim, Yoon-Ho and Kulik, Sergei P. and Shih, Yanhua},
	journal = {Phys. Rev. Lett.},
	volume = {86},
	issue = {7},
	pages = {1370--1373},
	numpages = {0},
	year = {2001},
	month = {2},
	publisher = {American Physical Society},
	doi = {10.1103/PhysRevLett.86.1370},
	url = {https://doi.org/10.1103/PhysRevLett.86.1370}
}

@article{Bayerbach2023,
	author = {Matthias J. Bayerbach  and Simone E. D’Aurelio  and Peter van Loock  and Stefanie Barz },
	title = {Bell-state measurement exceeding 50\% success probability with linear optics},
	journal = {Science Advances},
	volume = {9},
	number = {32},
	pages = {eadf4080},
	year = {2023},
	doi = {10.1126/sciadv.adf4080},
url = {https://doi.org/10.1126/sciadv.adf4080}
}

@article{Schuck2006,
	title = {Complete Deterministic Linear Optics Bell State Analysis},
	author = {Schuck, Carsten and Huber, Gerhard and Kurtsiefer, Christian and Weinfurter, Harald},
	journal = {Phys. Rev. Lett.},
	volume = {96},
	issue = {19},
	pages = {190501},
	numpages = {4},
	year = {2006},
	month = {5},
	publisher = {American Physical Society},
	doi = {10.1103/PhysRevLett.96.190501},
	url = {https://doi.org/10.1103/PhysRevLett.96.190501}
}

@Article{Barreiro2008,
	author={Barreiro, Julio T.
	and Wei, Tzu-Chieh
	and Kwiat, Paul G.},
	title={Beating the channel capacity limit for linear photonic superdense coding},
	journal={Nature Physics},
	year={2008},
	month={4},
	day={01},
	volume={4},
	number={4},
	pages={282-286},
	issn={1745-2481},
	doi={10.1038/nphys919},
	url={https://doi.org/10.1038/nphys919}
}

@article{Williams2017,
	title = {Superdense Coding over Optical Fiber Links with Complete Bell-State Measurements},
	author = {Williams, Brian P. and Sadlier, Ronald J. and Humble, Travis S.},
	journal = {Phys. Rev. Lett.},
	volume = {118},
	issue = {5},
	pages = {050501},
	numpages = {5},
	year = {2017},
	month = {2},
	publisher = {American Physical Society},
	doi = {10.1103/PhysRevLett.118.050501},
	url = {https://doi.org/10.1103/PhysRevLett.118.050501}
}

@article{Bakhshinezhad2024,
	title = {Scalable Entanglement Certification via Quantum Communication},
	author = {Bakhshinezhad, Pharnam and Mehboudi, Mohammad and Carceller, Carles Roch i and Tavakoli, Armin},
	journal = {PRX Quantum},
	volume = {5},
	issue = {2},
	pages = {020319},
	numpages = {16},
	year = {2024},
	month = {4},
	publisher = {American Physical Society},
	doi = {10.1103/PRXQuantum.5.020319},
	url = {https://doi.org/10.1103/PRXQuantum.5.020319}
}

@article{
	Hu2018,
	author = {Xiao-Min Hu  and Yu Guo  and Bi-Heng Liu  and Yun-Feng Huang  and Chuan-Feng Li  and Guang-Can Guo },
	title = {Beating the channel capacity limit for superdense coding with entangled ququarts},
	journal = {Science Advances},
	volume = {4},
	number = {7},
	pages = {eaat9304},
	year = {2018},
	doi = {10.1126/sciadv.aat9304},
url = {https://doi.org/10.1126/sciadv.aat9304}
}

@Article{Erhard2020,
	author={Erhard, Manuel
	and Krenn, Mario
	and Zeilinger, Anton},
	title={Advances in high-dimensional quantum entanglement},
	journal={Nature Reviews Physics},
	year={2020},
	month={7},
	day={01},
	volume={2},
	number={7},
	pages={365-381},
	issn={2522-5820},
	doi={10.1038/s42254-020-0193-5},
url = {https://doi.org/10.1038/s42254-020-0193-5}
}

@article{Hu2020d32,
  title = {Efficient Generation of High-Dimensional Entanglement through Multipath Down-Conversion},
  author = {Hu, Xiao-Min and Xing, Wen-Bo and Liu, Bi-Heng and Huang, Yun-Feng and Li, Chuan-Feng and Guo, Guang-Can and Erker, Paul and Huber, Marcus},
  journal = {Phys. Rev. Lett.},
  volume = {125},
  issue = {9},
  pages = {090503},
  numpages = {6},
  year = {2020},
  month = {8},
  publisher = {American Physical Society},
  doi = {10.1103/PhysRevLett.125.090503},
  url = {https://doi.org/10.1103/PhysRevLett.125.090503}
}

@article{PhysRevLett.125.090503,
  title = {Efficient Generation of High-Dimensional Entanglement through Multipath Down-Conversion},
  author = {Hu, Xiao-Min and Xing, Wen-Bo and Liu, Bi-Heng and Huang, Yun-Feng and Li, Chuan-Feng and Guo, Guang-Can and Erker, Paul and Huber, Marcus},
  journal = {Phys. Rev. Lett.},
  volume = {125},
  issue = {9},
  pages = {090503},
  numpages = {6},
  year = {2020},
  month = {8},
  publisher = {American Physical Society},
  doi = {10.1103/PhysRevLett.125.090503},
  url = {https://doi.org/10.1103/PhysRevLett.125.090503}
}

@article{Hu:20,
author = {Xiao-Min Hu and Wen-Bo Xing and Bi-Heng Liu and De-Yong He and Huan Cao and Yu Guo and Chao Zhang and Hao Zhang and Yun-Feng Huang and Chuan-Feng Li and Guang-Can Guo},
journal = {Optica},
number = {7},
pages = {738--743},
publisher = {Optica Publishing Group},
title = {Efficient distribution of high-dimensional entanglement through 11km fiber},
volume = {7},
month = {7},
year = {2020},
url = {https://opg.optica.org/optica/abstract.cfm?URI=optica-7-7-738},
doi = {10.1364/OPTICA.388773},
}

@article{PhysRevA.96.022317,
  title = {High-dimensional decoy-state quantum key distribution over multicore telecommunication fibers},
  author = {Ca\~nas, G. and Vera, N. and Cari\~ne, J. and Gonz\'alez, P. and Cardenas, J. and Connolly, P. W. R. and Przysiezna, A. and G\'omez, E. S. and Figueroa, M. and Vallone, G. and Villoresi, P. and da Silva, T. Ferreira and Xavier, G. B. and Lima, G.},
  journal = {Phys. Rev. A},
  volume = {96},
  issue = {2},
  pages = {022317},
  numpages = {8},
  year = {2017},
  month = {8},
  publisher = {American Physical Society},
  doi = {10.1103/PhysRevA.96.022317},
  url = {https://doi.org/10.1103/PhysRevA.96.022317}
}

@article{ding2017high,
       author = {{Ding}, Yunhong and {Bacco}, Davide and {Dalgaard}, Kjeld and {Cai}, Xinlun and {Zhou}, Xiaoqi and {Rottwitt}, Karsten and {Oxenl{\o}we}, Leif Katsuo},
        title = "{High-dimensional quantum key distribution based on multicore fiber using silicon photonic integrated circuits}",
      journal = {npj Quantum Information},
         year = 2017,
        month = {12},
       volume = {3},
       number = {1},
          eid = {25},
        pages = {25},
          doi = {10.1038/s41534-017-0026-2},
url = {https://doi.org/10.1038/s41534-017-0026-2}
}

@article{Schneeloch_2019,
doi = {10.1088/2040-8986/ab05a8},
url = {https://dx.doi.org/10.1088/2040-8986/ab05a8},
year = {2019},
month = {2},
publisher = {IOP Publishing},
volume = {21},
number = {4},
pages = {043501},
author = {James Schneeloch and Samuel H Knarr and Daniela F Bogorin and Mackenzie L Levangie and Christopher C Tison and Rebecca Frank and Gregory A Howland and Michael L Fanto and Paul M Alsing},
title = {Introduction to the absolute brightness and number statistics in spontaneous parametric down-conversion},
journal = {Journal of Optics}
}

@inproceedings{harrison2010introduction,
  title={Introduction to monte carlo simulation},
  author={Harrison, Robert L},
  booktitle={AIP conference proceedings},
  volume={1204},
  number={1},
  pages={17--21},
  year={2010},
  organization={American Institute of Physics},
url={https://ieeexplore.ieee.org/abstract/document/4736059}
}

@article{joy1991introduction,
  title={An introduction to Monte Carlo simulations},
  author={Joy, David C},
  journal={Scanning microscopy},
  volume={5},
  number={2},
  pages={4},
  year={1991},
url={https://digitalcommons.usu.edu/microscopy/vol5/iss2/4/}
}
\noindent
\subsection*{Acknowledgements}
This work was supported by the NSFC (No.~12374338, No.~12174367, No.~12204458, and No.~12350006), the Innovation Program for Quantum Science and Technology (No. 2021ZD0301200), the Fundamental Research Funds for the Central Universities, Anhui Provincial Natural Science Foundation (No. 2408085JX002),  Anhui Province Science and Technology Innovation Project (No. 202423r06050004), China Postdoctoral Science Foundation (2021M700138), China Postdoctoral for Innovative Talents (BX2021289). A.T is supported by the Wenner-Gren Foundations, by the Knut and Alice Wallenberg Foundation through the Wallenberg Center for Quantum Technology (WACQT) and the Swedish Research Council under Contract No.~2023-03498. J.P. is supported by NCCR-SwissMAP. This work was partially carried out at the USTC Center for Micro and Nanoscale Research and Fabrication.

\clearpage

\end{document}


\onecolumn{
\maketitle 
\vspace{-9mm}
\begin{center}
\begin{minipage}{1\textwidth}
\begin{center}
\textit{\textrm{
\textsuperscript{1} CAS Key Laboratory of Quantum Information, University of Science and Technology of China, Hefei 230026, China.
\\\textsuperscript{2} CAS Center For Excellence in Quantum Information and Quantum Physics, University of Science and Technology of China, Hefei, 230026, China.
\\\textsuperscript{3} Hefei National Laboratory, University of Science and Technology of China, Hefei 230088, China.
\\\textsuperscript{4} Department of Applied Physics, University of Geneva, 1211 Geneva, Switzerland.
\\\textsuperscript{5} Constructor Institute of Technology (CIT), Geneva, Switzerland.
\\\textsuperscript{6} Physics Department and NanoLund, Lund University, Box 118, 22100 Lund, Sweden.
\\\textsuperscript{7} These authors contributed equally: Chao Zhang, Jia-Le Miao.
\\\ ~~~Emails to: huxm@ustc.edu.cn; jef.pauwels@unige.ch; armin.tavakoli@teorfys.lu.se; bhliu@ustc.edu.cn}}
\end{center}
\end{minipage}
\end{center}

} 

\setcounter{figure}{0}
\makeatletter 
\renewcommand{\thefigure}{S\@arabic\c@figure}
\makeatother
 
\setcounter{table}{0}
\makeatletter  
\renewcommand{\thetable}{S\@arabic\c@table}
\makeatother

\setcounter{equation}{0}
\makeatletter 
\renewcommand{\theequation}{S\@arabic\c@equation}
\makeatother

\section{Optimal protocol with product measurements} \label{app:optimality}

We compute the probability distribution $p(b|x_1,x_2,y)$ associated with our protocol. 
In the main text, we defined $M_{b|1}=\sum_{a=0}^{n-1}\ketbra{t_{ab}}{t_{ab}}$ and $M_{b|2}=(F\otimes F^\dagger) M_{b|1}(F^\dagger \otimes F)$, where $\ket{t_{ab}}=\ket{a+b,a}$ and $F$ is the Fourier matrix defined by $F=\sum_{k,l=0}^{n-1} \omega^{kl}\ketbra{k}{l}$ where $\omega=e^{\frac{2\pi i}{n}}$. For the first measurement setting, we have the following probabilities:
\begin{align}\nonumber
	p(b | (x_1,x_2), 1) &=  \sum_{a=0}^{n-1} \abs{ \bra{t_{ab} }  Z^{x_2} X^{x_1} \otimes \mathbb{1} \ket{\phi_n}}^2 = \frac{1}{n} \sum_{a=0}^{n-1} \abs{ \sum_{k=0}^{n-1} \bra{a+b,a} Z^{x_2} X^{x_1} \otimes \mathbb{1} \ket{k,k} }^2 \\ \nonumber&= \frac{1}{n} \sum_{a=0}^{n-1} \abs{ \sum_{k=0}^{n-1} \bra{a+b,a} Z^{x_2} \otimes \mathbb{1} \ket{k+x_1, k} }^2 = \frac{1}{n}\sum_{a=0}^{n-1} \abs{ \sum_{k=0}^{n-1} \omega^{x_2(k+x_1)} \delta_{a+b,k+x_1} \delta_{a,k} }^2 \\ &= \frac{1}{n}\sum_{a=0}^{n-1} \abs{\omega^{x_2(a+x_1)}}^2 \delta_{b,x_1} = \delta_{b,x_1}\,.
\end{align}

Similarly, for the second measurement setting, we find:
\begin{align}\nonumber
	p(b | (x_1,x_2), 2) &=  \sum_{a=0}^{n-1} \abs{ \bra{t_{ab} }  F^\dagger Z^{x_2} X^{x_1} \otimes F \ket{\phi^+}}^2 = \frac{1}{n} \sum_{a=0}^{n-1} \abs{ \sum_{k=0}^{n-1} \bra{a+b,a} F^\dagger Z^{x_2} X^{x_1} \otimes F \ket{k,k} }^2 \\ \nonumber&= \frac{1}{n^2} \sum_{a=0}^{n-1} \abs{ \sum_{k,l,t=0}^{n-1} \omega^{-(a+b)l +kt}\bra{l,a}  Z^{x_2}  \otimes \mathbb{1} \ket{k+x_1,t} }^2 = \frac{1}{n^2} \sum_{a=0}^{n-1} \abs{ \sum_{k,l,t=0}^{n-1} \omega^{-(a+b)l +kt +x_2(k+x_1)} \delta_{l,k+x_1} \delta_{a,t} }^2 \\ &= \frac{1}{n^2} \sum_{a=0}^{n-1} \abs{\omega^{x_1x_2 -(a+b)x_1} \sum_{k=0}^{d-1}  \omega^{k(x_2-b)}}^2 = \delta_{x_2,b} \, .   
\end{align}

\section{Bound for Schmidt number certification} \label{app:schmidtnumberbound}
Let Alice's encoding channels be defined as $\Lambda_{x}:\mathcal{L}(\mathbb{C}^N)\rightarrow \mathcal{L}(\mathbb{C}^n)$. These are completely positive trace-preserving maps that transform an $N$-dimensional system into an $n$-dimensional system, and we denote the set of such maps by $\mathcal{T}_n$. Here, $N$ is arbitrary while $n$ is given. Denoting the shared state by $\psi_{AB}$, the total state received by Bob becomes $\tau_x=(\Lambda_x\otimes \mathbb{1})[\psi_{AB}]$. Bob performs joint measurements $\{M_{b|y}\}$ on this two-particle state. The probability distribution is then given by $p(b|x,y)=\tr\left(\tau_x M_{b|y}\right)$. The success rate in stochastic communication can then be expressed as
\begin{equation}
\mathcal{S}=\frac{1}{2n^2}\sum_{x_1,x_2,y} \tr\left(M_{x_y|y} (\Lambda_{x_1x_2}\otimes \mathbb{1})[\psi]\right).
\end{equation}

In general, we permit $\psi$ to have Schmidt number at most $d$. The Schmidt number is defined as
\begin{align}\label{schmidt}
	\text{SN}(\psi_{AB})\equiv \min_{\{p_\lambda\},\{\varphi_\lambda\}}  \Big\{&d_\text{max}: \quad  \psi_{AB}=\sum_\lambda p_\lambda \ketbra{\varphi_\lambda}
	\quad \text{and} \quad d_\text{max}=\max_\lambda \text{SR}(\varphi_\lambda)\Big\},
\end{align}
where $\{p_\lambda\}$ is a probability distribution, $\varphi_\lambda$ are pure states and $\text{SR}$ denotes the Schmidt rank of a pure state, which is defined as $\text{SR}(\varphi)=\rank(\varphi_A)$, where $\varphi_A=\tr_B(\varphi_{AB})$. The Schmidt number of $\psi_{AB}$ is the largest Schmidt rank appearing in the least dimension-costly ensemble decomposition of the state $\psi_{AB}$. In the case of pure states, the Schmidt number reduces to the Schmidt rank.

Thus, we are interested in the following optimisation problem.

\begin{align}
	\mathcal{S}_d\equiv \max \quad &\frac{1}{2n^2}\sum_{x_1,x_2,y} \tr\left(M_{x_y|y} (\Lambda_{x_1x_2}\otimes \mathbb{1})[\psi]\right)\\
	& \text{s.t}\quad M_{b|y}\succeq 0, \quad \sum_b M_{b|y}=\mathbb{1}, \quad \psi\succeq 0, \quad \tr(\psi)=1,\quad \text{SN}(\psi)\leq d, \quad \Lambda_x\in \mathcal{T}_n,
\end{align}
where $\psi\in \mathcal{L}(\mathcal{C}^N\otimes \mathcal{C}^N)$ for any $N$. While this optimisation is hard to solve, we will derive an upper bound on $\mathcal{S}_d$ obtained via a simple relaxation of the scenario.

Firstly, since $\mathcal{S}$ is linear in $\psi$, one can easily show that $\psi$ can w.l.g be selected as pure. Then the Schmidt number is equivalent to the Schmidt rank, i.e. $\text{SN}(\psi)=\text{SR}(\psi)$. Under local unitaries (which can always be moved into the labs of Alice and Bob) the pure state $\psi$ can be transformed into the space $\mathbb{C}^d\otimes \mathbb{C}^d$. Hence, we wish to maximize $\mathcal{S}$ over all $d\times d$ pure states $\psi$. Notice now that the state $\tau_x=(\Lambda_x\otimes \mathbb{1})[\psi]$ becomes a bipartite state on $\mathbb{C}^n\otimes \mathbb{C}^d$. It is also subject to the no-signaling constraint $\tr_A(\tau_x) = \tr_A (\tau)$. 

Our relaxation consists in simply ignoring the no-signaling constraint. Thus, we consider the strict superset of states $\{\rho_x\}$ consisting of arbitrary quantum states of dimension $nd$. We call the figure of merit in this relaxed scenario $\mathcal{S}'$. It reads
\begin{equation}
\mathcal{S}'=\frac{1}{2n^2}\sum_{x_1,x_2=0}^{n-1} \Tr\left[\rho_x (P_{x_1}+Q_{x_2})\right],
\end{equation}
where $P$ and $Q$ are Bob's two measurements. We need to put a bound on $\mathcal{S}'$ valid for arbitrary $\{\rho_x\}$, $\{P_b\}$ and $\{Q_b\}$. To achieve this, we follow the methods of Refs.\cite{PhysRevA.99.032316} and \cite{Farkas2024}. By selecting the optimal pure state $\rho_x$, we obtain 
\begin{align}
\mathcal{S}'\leq \frac{1}{2n^2}\sum_{x_1,x_2=0}^{n-1} \norm{P_{x_1}+Q_{x_2}}_\infty.
\end{align}
The norm corresponds to the largest modulus eigenvalue. This is saturated by choosing $\rho_x$ as the pure state corresponding to the best eigenvector of $P_{x_1}+Q_{x_2}$. Next we use the operator relation of Kittaneh \cite{Kittaneh1997} used in \cite{PhysRevA.99.032316}, namely that for any $A,B\succeq 0$ it holds that 
\begin{equation}
\norm{A+B}\leq \max\left(\norm{A},\norm{B}\right)+\norm{\sqrt{A}\sqrt{B}}.
\end{equation}
Applying this gives
\begin{equation}
\mathcal{S'}\leq  \frac{1}{2n^2}\sum_{x_1,x_2=0}^{n-1} \left(\max\left(\norm{P_{x_1}},\norm{Q_{x_2}}\right)+\norm{\sqrt{P_{x_1}}\sqrt{Q_{x_2}}}\right).
\end{equation}
Since $P$ and $Q$ are measurements, it follows that $\max\left(\norm{P_{x_1}},\norm{Q_{x_2}}\right)\leq 1$. Hence,
\begin{equation}
\mathcal{S'}\leq  \frac{1}{2}+\frac{1}{2n^2}\sum_{x_1,x_2=0}^{n-1}\norm{\sqrt{P_{x_1}}\sqrt{Q_{x_2}}}.
\end{equation}
Then we can use that the Schatten norms are ordered, namely that $\lVert \cdot \rVert_a\leq\lVert \cdot \rVert_b $ when $b\leq a$. We bound the infinity norm with the two-norm $\lVert A \rVert_2=\sqrt{\Tr\left(AA^\dagger\right)}$. Then,
\begin{equation}
\mathcal{S'}\leq  \frac{1}{2}+\frac{1}{2n^2}\sum_{x_1,x_2=0}^{n-1} \sqrt{\Tr\left(P_{x_1}Q_{x_2}\right)}.
\end{equation}
Then we use the concavity-inequality of the square-root, namely for $x_i\geq 0$ it holds that $\sum_{i=1}^m \sqrt{x_i}\leq \sqrt{m}\sqrt{\sum_i x_i}$ with equality if and only if all $x_i$ are equal. It gives
\begin{align}
\mathcal{S}\leq \mathcal{S'}&\leq  \frac{1}{2}+\frac{1}{2n}\sqrt{\sum_{x_1,x_2=0}^{n-1}\Tr\left(P_{x_1}Q_{x_2}\right)}\\
&= \frac{1}{2}+\frac{1}{2n}\sqrt{\Tr(\mathbb{1}_{nd})}=\frac{1}{2}\left(1+\sqrt{\frac{d}{n}}\right).
\end{align}
This is an analytical bound whose violation implies a Schmidt number larger than $d$. 

The bound is tight for $d=1$ and $d=n$. However, as expected, the bound is not tight for general $d$. To probe its accuracy, we considered a few test cases $(n,d)=(8,2),(8,3),(8,4)$ and numerically optimized. We find that the bound is accurate up to $1-2 \%$.

\section{Improving the robustness of Schmidt number detection} \label{app:prime}

Following \cite{Bakhshinezhad2024}, one can construct a communication protocol for prime channel dimensions $n$ and $m\in\{2,\ldots,n+1\}$ bases. We show that such protocols can be used to enhance the noise-robustness of Schmidt number  certification. In prime dimensions, a complete set of $n+1$ mutually unbiased bases (MUBs) can be constructed as follows. The first $d$ bases are given by
\begin{equation}
\ket*{e^{(y)}_l}=\frac{1}{\sqrt{n}}\sum_{k=0}^{n-1} \omega^{k(l+yk)}\ket{k},
\end{equation}
where $l\in\{0,\ldots,n-1\}$ labels the basis elements and $y\in\{0,\ldots,n-1\}$ the basis. We have defined $\omega=e^{2\pi i/n}$. The final MUB, $y=n$, can be chosen as the computational basis $\ket*{e^{(n)}_l}=\ket{l}$. 

As in the protocol in the main text, we choose as encoding the $n^2$ dense coding unitaries, $U_x=X^{x_1}Z^{x_2}$, on the maximally entangled state $\ket{\phi^+}$. This leads to the total state $\ket{\phi_x}=U_x\otimes \mathbb{1} \ket{\phi^+}$ in the lab of the receiver, who in turn projects the first and second subsystems onto the $y$'th MUB and its conjugate respectively, leading to the outcome $l_1$ and $l_2$. 

The resulting probability distribution is given by
\begin{equation}
p(l_1,l_2|x,y)=| \braket*{\phi_x}{e^{(y)}_{l_1},e^{(y)*}_{l_2}}|^2,
\end{equation}
where the star denotes complex conjugation and $\ket{\phi_x}=\frac{1}{\sqrt{n}}\sum_k \omega^{x_2 k}\ket{k+x_1,k}$. For $y=0,\ldots,n-1$ these evaluate to
	
\begin{align}
p(l_1,l_2|x,y)&=\frac{1}{n^3} \bigg|\sum_{s,k,k'}\omega^{-x_2 s+k(l_1+yk)-k'(l_2+yk')}\underbrace{\braket{s+x_1,s}{k,k'}}_{\Rightarrow s=k' \text{ and } k'=k-x_1}\bigg|^2=\frac{1}{n^3}\left|\omega^{x_1x_2+x_1l_2-yx_1^2} \sum_k \omega^{k(l_1-l_2-x_2+2yx_1)}\right|^2\\
& = \frac{1}{n^3}\bigg|\underbrace{\sum_k \omega^{k(l_1-l_2-x_2+2yx_1)}}_{=n \delta_{l_1-l_2,x_2-2yx_1}}\bigg|^2=\begin{cases}
\frac{1}{n} & \text{ if } l_1-l_2=x_2-2yx_1\\
0  & \text{otherwise}
\end{cases}.
\end{align}

Thus, for a single $n$-valued outcome $l\equiv l_1-l_2$ this leads to the winning conditions
\begin{equation}
l\equiv l_1-l_2=\begin{cases}
x_2-2yx_1 & \text{if } y=0,\ldots,n-1\\
x_1 & \text{if } y=n
\end{cases}.
\end{equation}
Note that the last line (computational basis) follows from our previous consideration of $Z\otimes Z$ measurement for the stochastic communication protocol considered in the main text. This leads to the following linear game function \cite{Bakhshinezhad2024}.

\begin{equation}\label{witness}
\mathcal{R}=\frac{1}{md^2} \sum_{x_1,x_2}\left(p(l=x_1|x,y=d)+ \sum_{y=0}^{m-2} p(l=x_2-2yx_1|x,y)\right) \,.
\end{equation}
By construction, we achieve $\mathcal{R}=1$ by the above protocol.

Like the stochastic communication protocol of the main text, the success of the protocol crucially hinges on the parties sharing high-dimensional entanglement. We determine bounds on $\mathcal{R}$ valid for any protocol based on entanglement of Schmidt number $d<n$. We address this in the same way as for the stochastic communication task in the main text, namely by relaxing the problem to a prepare-and-measure scenario in which arbitrary states of dimension $dn$ may be communicated. Correlations arising in such scenarios can be bounded using semidefinite relaxation methods \cite{SDPreview}. We use the relaxation hierarchy of \cite{Pauwels2022b} to bound $\mathcal{R}$. These programs become computationally expensive very fast because of the large number of states and measurements. To remedy this, we use a low relaxation level corresponding to all monomials of length one combined with all monomials of length two corresponding to the products $\rho_xM_{b|y}$ that appear in the objective function $\mathcal{R}$.  The size of the moment matrix becomes roughly
\begin{equation}
\dim\Gamma\sim 2+n^2+mn+mn^2.
\end{equation}
For example, for $n=7$ and $m=n+1$ bases, the moment matrix is of size $498$, which is manageable. In Table~\ref{Tab:primemub} we list the resulting bounds on $\mathcal{R}$ for $n=3,5,7$, for any number of MUBs, $m=2,\ldots,n+1$, and any Schmidt number $d=1,\ldots,n-1$. These bounds are not expected to be optimal. The numbers marked in blue enable stronger Schmidt number detection than does the  analytical bound derived for the stochastic communication task when applied to the isotropic state
\begin{equation}
    v\ketbra{\phi^+}+\frac{1-v}{d^2}\mathbb{1}.
\end{equation}

\begin{table}[h]
        \centering
	\begin{tabular}{|c|c|c|c|c|c|c|c|}
		\hline
		 &  2                                                                                       &   3                                                                                     & 4                                                                                                                & 5                                                                                                                & 6                                                                                                                & 7                                                                                                                & 8                                                                                                                \\ \hline
		3                        & \begin{tabular}[c]{@{}c@{}}0.8024\\ 0.9692\end{tabular}                                                          & \begin{tabular}[c]{@{}c@{}}\blue{0.7182}\\ 0.9285\end{tabular}                                                          & \begin{tabular}[c]{@{}c@{}}\blue{2/3}\\ \blue{0.8604}\end{tabular}                                                             & -                                                                                                                & -                                                                                                                & -                                                                                                                & -                                                                                                                \\ \hline
		5                        & \begin{tabular}[c]{@{}c@{}}0.7553\\ 0.9190\\ 0.9761\\ 0.9958\end{tabular}                                        & \begin{tabular}[c]{@{}c@{}} \blue{0.6618}\\ 0.8562\\ 0.9491\\ 0.9897\end{tabular}                                        & \begin{tabular}[c]{@{}c@{}} \blue{3/5}\\ \blue{0.7971}\\ 0.9160\\ 0.9801\end{tabular}                                           & \begin{tabular}[c]{@{}c@{}}\blue{0.5578}\\ \blue{0.7367}\\ \blue{0.8690}\\ 0.9618\end{tabular}                                        & \begin{tabular}[c]{@{}c@{}}\blue{0.5266}\\ \blue{0.6899}\\ \blue{0.8110}\\ \blue{0.9118}\end{tabular}                                        & -                                                                                                                & -                                                                                                                \\ \hline
		\multicolumn{1}{|l|}{7}  & \multicolumn{1}{l|}{\begin{tabular}[c]{@{}l@{}}\blue{0.7318}\\ 0.8903\\ 0.9530\\ 0.9810\\ 0.9936\\ 0.9988\end{tabular}} & \multicolumn{1}{l|}{\begin{tabular}[c]{@{}l@{}}\blue{0.6352}\\ 0.8181\\ 0.9109\\ 0.9605\\ 0.9859\\ 0.9971\end{tabular}} & \multicolumn{1}{l|}{\begin{tabular}[c]{@{}l@{}}\blue{0.5714}\\ \blue{0.7597}\\ 0.8703\\ 0.9375\\ 0.9760\\ 0.9948\end{tabular}} & \multicolumn{1}{l|}{\begin{tabular}[c]{@{}l@{}}\blue{0.5262}\\ \blue{0.7066}\\ 0.8285\\ 0.9106\\ 0.9630\\ 0.9914\end{tabular}} & \multicolumn{1}{l|}{\begin{tabular}[c]{@{}l@{}}\blue{0.4928}\\ \blue{0.6579}\\ \blue{0.7817}\\ \blue{0.8768}\\ 0.9443\\ 0.9857\end{tabular}} & \multicolumn{1}{l|}{\begin{tabular}[c]{@{}l@{}}\blue{0.4668}\\ \blue{0.6197}\\ \blue{0.7343}\\ \blue{0.8301}\\\blue{ 0.9129}\\ 0.9737\end{tabular}} & \multicolumn{1}{l|}{\begin{tabular}[c]{@{}l@{}}\blue{0.4459}\\ \blue{0.5889}\\ \blue{0.6961}\\ \blue{0.7857}\\ \blue{0.8643}\\ \blue{0.9350}\end{tabular}} \\ \hline
	\end{tabular}
\caption{Rows correspond to $n=3,5,7$. Columns correspond to the number of MUBs, $m=2,\ldots,n+1$. Each box lists the bounds on $\mathcal{R}$ for Schmidt numbers $1,\ldots,n-1$.}\label{Tab:primemub}
\end{table}

\section{Construction of 8-dimensional encodings $Z^{x_2}$ and $X^{x_1}$}\label{app:constructionofu}

\begin{figure}[h]
    \centering
    \includegraphics[width=0.5\linewidth]{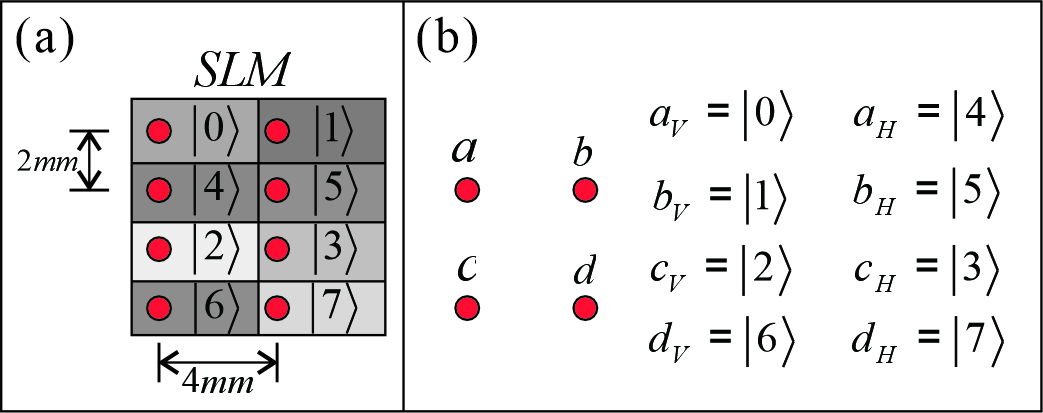}
    \caption{
    (a) Along the direction of light propagation, the encoding scheme for the subspaces of the 8-dimensional entangled state  $|\Phi\rangle$  is illustrated, along with the corresponding segmented regions on the SLM.
    (b) After performing the operation  $Z^{x_2}$, the entangled state  $|\Phi\rangle $ is encoded in the path-polarization degrees of freedom (DOFs). The subscript $H (V)$ indicates that the subspace is in the horizontal (vertical) polarization state of the corresponding path.}
    \label{code}
\end{figure}

As illustrated in Fig.~\ref{code}(a), the 8-dimensional entangled state
\begin{equation}
   \left|\Phi\right\rangle=\frac{1}{\sqrt{8}} \sum_{k=0}^{7}|k_H\rangle_A |k_V\rangle_B 
\end{equation}
generated via the SPDC process is encoded into the path degree of freedom (DOF), forming a $4\times2$ matrix of rays with a horizontal spacing of 4 mm and a vertical spacing of 2 mm. On Alice's side, the phase of each subspace can be independently controlled by adjusting the grayscale values of the eight regions segmented by the spatial light modulator (SLM). 

When performing 
the operation \( Z^{x_2} \),  \( x_2 \in \{0, 1, \ldots, 7\} \), on the initially entangled state $|\Phi\rangle$, a phase shift of \( e^{\frac{\pi i}{4} k x_2} \) is introduced in the subspaces \( |k\rangle_A \), as detailed below:
\begin{equation}
\begin{array}{lllllllll}
\mathbf{subspace} & \quad Z^0 &\;\,\quad Z^1 &\;\,\quad Z^2 &\;\,\quad Z^3 &\;\,\quad Z^4 &\;\,\quad Z^5 &\;\,\quad Z^6 &\;\,\quad Z^7  \\
\begin{array}{l}
|0\rangle_A\\
|1\rangle_A\\
|2\rangle_A\\
|3\rangle_A\\
|4\rangle_A\\
|5\rangle_A\\
|6\rangle_A\\
|7\rangle_A\\
        \end{array} & \left(\begin{array}{l}
1\\
1\\
1\\
1\\
1\\
1\\
1\\
1\\
        \end{array}\right) & \left(\begin{array}{l}
\;\;\,1\\
e^{\frac{\pi i}{4}}\\
\;\;\,i\\
e^{\frac{3\pi i}{4}}\\
-1\\
e^{\frac{5\pi i}{4}}\\
-i\\
e^{\frac{7\pi i}{4}}\\
        \end{array}\right) & \left(\begin{array}{l}
\;\;\,1\\
\;\;\,i\\
-1\\
-i\\
\;\;\,1\\
\;\;\,i\\
-1\\
-i\\
        \end{array}\right) & \left(\begin{array}{l}
\;\;\,1\\
e^{\frac{3\pi i}{4}}\\
-i\\
e^{\frac{\pi i}{4}}\\
-1\\
e^{\frac{7\pi i}{4}}\\
\;\;\,i\\
e^{\frac{5\pi i}{4}}\\
        \end{array}\right) & \left(\begin{array}{l}
\;\;\,1\\
-1\\
\;\;\,1\\
-1\\
\;\;\,1\\
-1\\
\;\;\,1\\
-1\\
        \end{array}\right) & \left(\begin{array}{l}
\;\;\,1\\
e^{\frac{5\pi i}{4}}\\
\;\;\,i\\
e^{\frac{7\pi i}{4}}\\
-1\\
e^{\frac{\pi i}{4}}\\
-i\\
e^{\frac{7\pi i}{4}}\\
        \end{array}\right) & \left(\begin{array}{l}
\;\;\,1\\
-i\\
-1\\
\;\;\,i\\
\;\;\,1\\
-i\\
-1\\
\;\;\,i\\
        \end{array}\right) & \left(\begin{array}{l}
\;\;\,1\\
e^{\frac{7\pi i}{4}}\\
-i\\
e^{\frac{5\pi i}{4}}\\
-1\\
e^{\frac{3\pi i}{4}}\\
\;\;\,i\\
e^{\frac{\pi i}{4}}\\
        \end{array}\right).
\end{array}   \label{Z_operation}
\end{equation}
Therefore, based on the prepared initial state $|\Phi\rangle$, Alice can independently adjust the grayscale values of each region of the SLM to introduce the specific phase in Eq.~(\ref{Z_operation}) in each subspace. \\

After operation $Z^{x_2}$, as described in the experimental setup in Fig.~\ref{setup}, HWPA2 is then used to adjust the polarization of each path. Meanwhile, BD4 shifts the paths in rows 1 and 3 downward by 2 mm, forming a \(2 \times 2\) matrix with two polarization states for each path (`a, b, c, d'). The detailed encoding strategy is presented in Fig.~\ref{code}(b). In the following section, we describe in detail the construction of the operation $X^{x_1}$. \\

\begin{figure}[htb]
    \centering
    \includegraphics[width=1.0\linewidth]{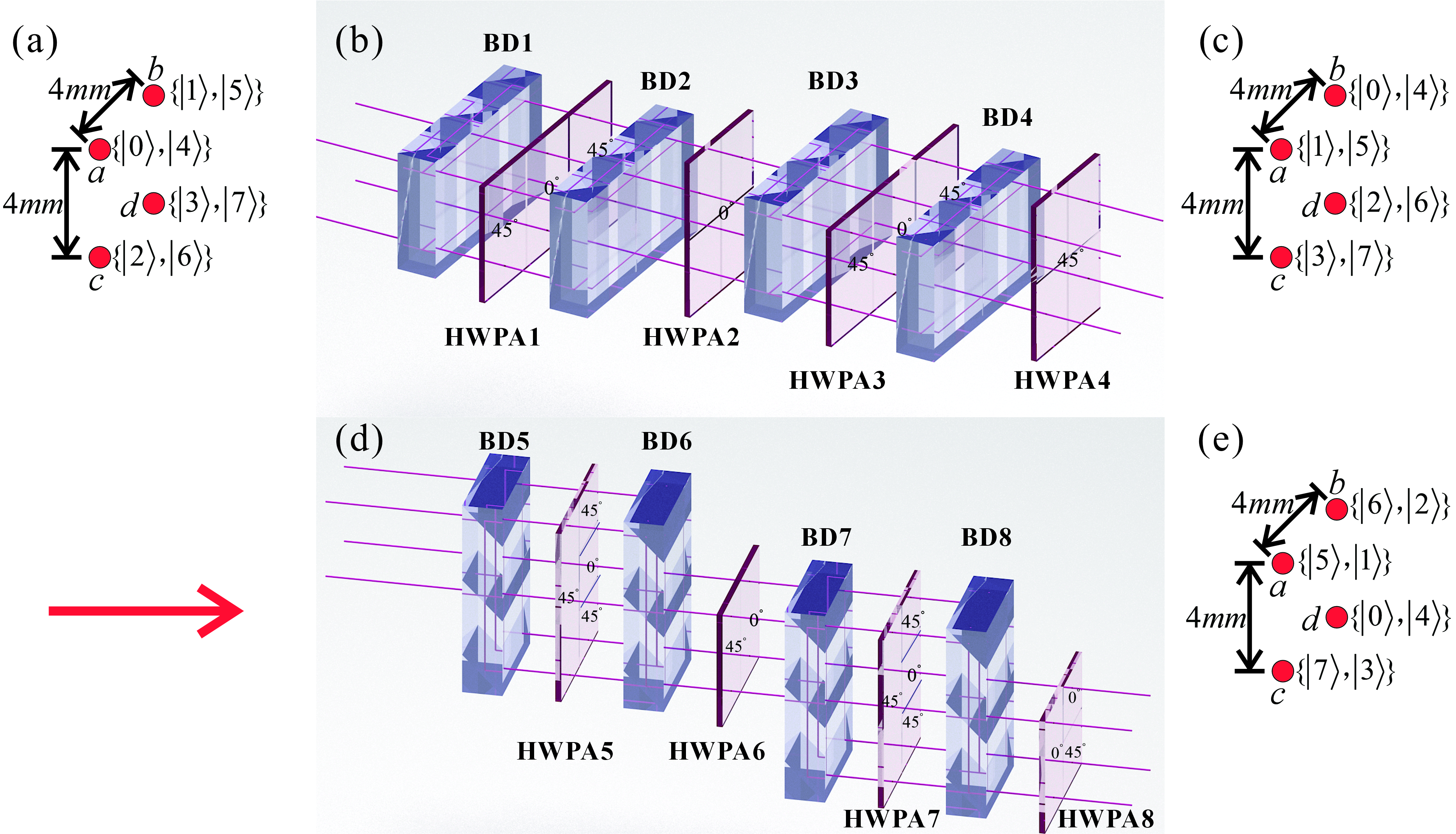}
    \caption{Construction of shift operation $X^{3}$. (a) The subspaces distribution of Alice's photon in each path, following the encoding method described in Fig.~\ref{code}(b). (b) Permutation of subspaces between left and right paths. BD1-BD4 reflect vertically polarized photons to the left as parallel beams with a spacing of 4 mm along the propagation direction, while horizontally polarized photons transmit through them. (c) Subspace distribution after left-right permutation, which serves as the input for the next part to achieve operation $X^{3}$. (d) Permutation of subspaces between upper and lower paths. BD5-BD8 reflect horizontally polarized photons downward as parallel beams with a spacing of 4 mm along the propagation direction, while vertically polarized photons transmit through them. (e) Output of $X^{3}$ operation. Each initial subspace of input photon is transformed into different path and polarization, i.e. another subspace. For example, subspace $\ket{7}$ for Alice's photon is in path $d$ with horizontally polarization originally. After operation $X^3$, photons initially in $\ket{7}$ are now in path $c$ with vertical polarization, which mean the encoding of the subspace of Alice's photon has changed from $\ket{7}$ to $\ket{2}$, as shown in Fig.~\ref{code} (b). Note that HWPAs consist of half-wave plates at different angles, and they can be replaced to perform other shift operations. }
    \label{operation}
\end{figure}

Note that the subspace is encoded in the path DOF with planar extension, enabling shift operations in both the horizontal and vertical directions, which significantly accelerates high-dimensional operations. The construction process of the shift operation $X^3$ is illustrated in Fig.~\ref{operation}. The initial distribution of Alice's subspaces is shown in Fig.~\ref{operation} (a). In Fig.\ref{operation} (b), photons with horizontal polarization are transmitted through BD1-BD4, while photons with vertical polarization are reflected to the left as parallel beams with a spacing of 4 mm along the direction of light propagation. By properly designing the angles of HWPA1-HWP4, the polarization of the subspaces on each path can be controlled to complete the shift operation in the horizontal direction. The resulting subspace distribution of Alice's photons is shown in Fig.~\ref{operation} (c). Subsequently, as shown in Fig.~\ref{operation} (d), photons with vertical polarization are transmitted through BD5-BD8, while photons with horizontal polarization are reflected downward as parallel beams with a spacing of 4 mm along the direction of light propagation. Similarly, by properly designing the angles of HWPA4-HWPA8, the polarization of the subspaces on each path can be controlled to complete the shift operation in the vertical direction. The final distribution of Alice's subspaces is shown in Fig.~\ref{operation} (e), and compared to the initial subspace in Fig.~\ref{operation} (a), the shift operation $X^3$ on Alice photons is successfully implemented. Taking Alice's photon in $\ket{7}_A$ as an example, the photon is initially located in path $d$ with horizontal polarization, according to the encoding scheme. Then, Alice performs the shift operation $X^3$ on the photon, after which the photon initially encoded in $\ket{7}_A$ is now shifted to path $c$ with vertical polarization. Recall the encoding scheme of Fig.~\ref{code} (b), the photon is encoded in subspace $\ket{2}_A$ now. The operation for photons in remaining subspaces is similar to $\ket{7}_A$, so after the operation in Fig.~\ref{operation}, the original state 
\begin{equation}
    |\Phi\rangle_{AB}=\frac{1}{\sqrt{8}}(\ket{00}+\ket{11}+\ket{22}+\ket{33}+\ket{44}+\ket{55}+\ket{66}+\ket{77})  \notag
\end{equation}
will be transformed into 
\begin{equation}
    |\Phi\rangle_{AB}=\frac{1}{\sqrt{8}}(\ket{30}+\ket{41}+\ket{52}+\ket{63}+\ket{74}+\ket{05}+\ket{16}+\ket{27}). \notag
\end{equation}
We have preformed the complete operation $X^3$ on Alice's photons.\\

\begin{figure}[h]
    \centering
    \includegraphics[width=1\linewidth]{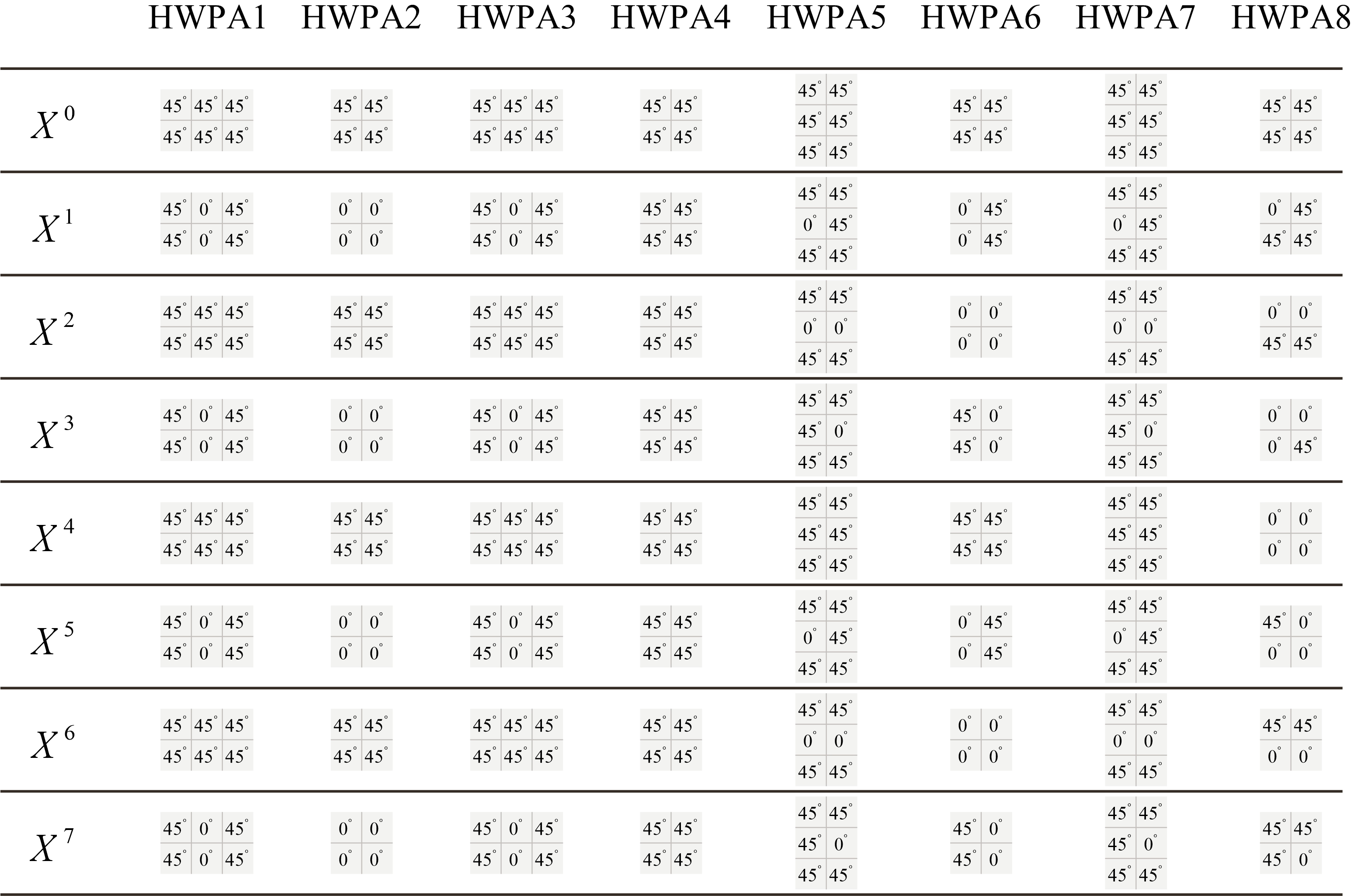}
    \caption{HWPAs constituting $X^{x_1}$ operation. By selecting a set of HWPAs, Alice can perform different shift operation to encode her photon.}
    \label{HWPAs}
\end{figure}

Similarly, for the remaining shift operation $X^{x_1}$ to be implemented, the corresponding angles must be redesigned for HWPA1-HWP8. In Fig.~\ref{HWPAs}, we provide the detailed angles for each HWPAs required to implement all shift operations. Therefore, we only need to replace HWPA1-HWPA8 to achieve the required shift operation $X^{x_1}$, $x_1\in\{0,1,\dots7\}$.\\

\section{Construction of 8-dimensional measurement for $X$ and $Z$}\label{app:high-dimensional measurement}

In the previous section, we provided a detailed explanation of how Alice constructs high-dimensional clock operations $Z^{x_2}$ and shift operations $X^{x_1}$, where $x_1, x_2 \in \{0,1,\dots,7\}$, to achieve dense encoding on her photons. After encoding her quantum state, Alice transmits it to Bob through a quantum communication channel with dimension $n=8$. Upon receiving the state, Bob performs measurements on the two photons based on his input $y \in \{1,2\}$: he applies the measurement $Z\otimes Z$ if $y=0$ and the measurement $X\otimes X$ if $y=1$. Bob constructs the measurements for the two received photons in the same manner. In the following section, we will provide a detailed description of how to construct the high-dimensional measurements for $X$, with eigenstates $\{|x_i\rangle=\sum_{k=0}^{n-1} e^{\frac{2 \pi i}{n} k l}|k\rangle\}$, and for $Z$, with eigenstates $\{|z_i\rangle=|k\rangle\}$.\\

\begin{figure}[h]
    \centering
    \includegraphics[width=0.7\linewidth]{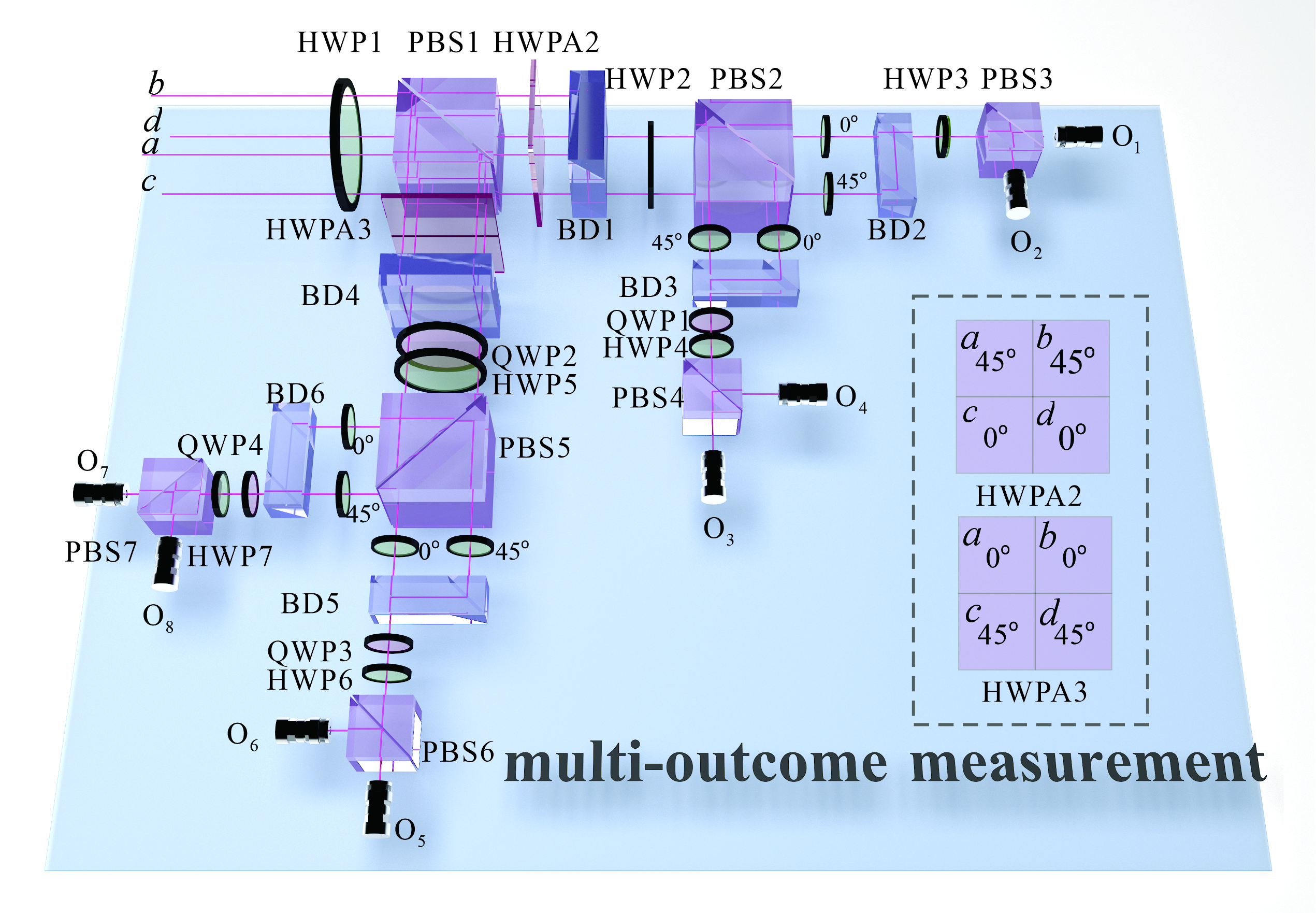 }
    \caption{Construction of high-dimensional multi-outcome measurement. BD2 and BD5 refracts vertically polarized photon in the horizontal direction by 4~mm. BD3, BD4, BD6 and BD7 refract horizontally polarized photons in the vertical direction by 4~mm. The information is encoded in the path DOF with $2\times 2$ array, labeled by $a$, $b$, $c$ and $d$, and polarization DOF. Each coupler $O_i$ collects the results of projection measurement $|x_i\rangle$ or $|z_i\rangle$. Bob constructs the same measurement setup for two photons, and by adjusting the angle of HWP1-HWP7 and QWP1-QWP4, he can perform $Z\otimes Z$ or $X\otimes X$ measurement with 8-outcome.}
    \label{measurementfig}
\end{figure}

Encoding subspaces on rectangular regions within a plate using path-polarization DOFs also offers significant advantages for constructing high-dimensional measurements. Prior to measurement, the subspace encoding method is illustrated in Fig.~\ref{code}(b). HWP1, combined with PBS1, constructs projection measurements for the two-dimensional subspaces $\{|0\rangle, |4\rangle\}$, $\{|1\rangle, |5\rangle\}$, $\{|2\rangle, |6\rangle\}$, and $\{|3\rangle, |7\rangle\}$ simultaneously. Next, HWP2 (HWP3 and QWP1), combined with PBS2 (PBS3), constructs projection measurements for the subspaces $\{|0\rangle, |4\rangle, |2\rangle, |6\rangle\}$ and $\{|1\rangle, |5\rangle, |3\rangle, |7\rangle\}$ simultaneously. Finally, by using the HWP and PBS in front of the coupler, a complete eight-dimensional projection measurement can be constructed. This encoding method allows multiple subspaces to be coherently processed simultaneously within the same optical components, significantly accelerating high-dimensional measurements. Furthermore, by replacing the waveplates in front of each PBS with a combination of \{QWP, HWP, QWP\}, any 8-dimensional measurement can be fully realized. The waveplates positioned in the measurement section of Fig.~\ref{measurementfig} are sufficient to construct the target measurements for $Z$ and $X$ in our experiment.\\

When Bob's input is $y=0$ ($y=1$), he performs the measurement $Z$ ($X$) on the two received photons independently. Bob set HWP1-HWP7 and QWP1-QWP4 to the angles specified in Table \ref{angle}, allowing the construction of $Z$ and $X$ measurements.\\

\begin{table}[h]
    \centering
    \renewcommand\arraystretch{1.3}
    \begin{tabular}{c|c|c|c|c|c|c|c|c|c|c|c}
    \hline
       Measurement & HWP1 & HWP2 & HWP3 & HWP4 & HWP5 & HWP6 & HWP7 & QWP1 & QWP2 & QWP3 & QWP4  \\
       \hline
      $Z$ &  $45^\circ$ & $0^\circ$ & $45^\circ$ & $45^\circ$ & $0^\circ$ & $45^\circ$ & $45^\circ$ & $0^\circ$ & $0^\circ$ & $0^\circ$ & $0^\circ$ \\
      \hline
      $X$ &  $22.5^\circ$ & $22.5^\circ$ & $22.5^\circ$ & $45^\circ$ & $45^\circ$ & $33.75^\circ$ & $11.25^\circ$ & $0^\circ$ & $0^\circ$ & $45^\circ$ & $45^\circ$ \\
      \hline
    \end{tabular}
    \caption{The angle at which the waveplates are set when constructing the measurements $Z$ and $X$.}
    \label{angle}
\end{table}

According to the angles in Table \ref{angle}, the projection measurements corresponding to outcome $O_1$ through $O_8$ are constructed in the computational basis when performing the $Z$ measurement:
\begin{equation}
\begin{array}{llllllll}
O_1=\left(\begin{array}{l}
1\\
0\\
0\\
0\\
0\\
0\\
0\\
0\\
        \end{array}\right), & O_2=\left(\begin{array}{l}
0\\
1\\
0\\
0\\
0\\
0\\
0\\
0\\
        \end{array}\right),& O_3=\left(\begin{array}{l}
0\\
0\\
1\\
0\\
0\\
0\\
0\\
0\\
        \end{array}\right),& O_4=\left(\begin{array}{l}
0\\
0\\
0\\
1\\
0\\
0\\
0\\
0\\
        \end{array}\right),\\ O_5=\left(\begin{array}{l}
0\\
0\\
0\\
0\\
1\\
0\\
0\\
0\\
        \end{array}\right),& O_6=\left(\begin{array}{l}
0\\
0\\
0\\
0\\
0\\
1\\
0\\
0\\
        \end{array}\right),& O_7=\left(\begin{array}{l}
0\\
0\\
0\\
0\\
0\\
0\\
1\\
0\\
        \end{array}\right), & O_8=\left(\begin{array}{l}
0\\
0\\
0\\
0\\
0\\
0\\
0\\
1\\
        \end{array}\right).
\end{array}   \label{Z_eigenstate}
\end{equation}

\clearpage
For the $X$ measurement, these projection measurements correspond to the Fourier basis:
\begin{equation}
\begin{array}{llll}
O_1=\left(\begin{array}{l}
1\\
1\\
1\\
1\\
1\\
1\\
1\\
1\\
        \end{array}\right), & O_2=\left(\begin{array}{l}
\;\;\,1\\
-1\\
\;\;\,1\\
-1\\
\;\;\,1\\
-1\\
\;\;\,1\\
-1\\
        \end{array}\right), & O_3=\left(\begin{array}{l}
\;\;\,1\\
\;\;\,i\\
-1\\
-i\\
\;\;\,1\\
\;\;\,i\\
-1\\
-i\\
        \end{array}\right), & O_4=\left(\begin{array}{l}
\;\;\,1\\
-i\\
-1\\
\;\;\,i\\
\;\;\,1\\
-i\\
-1\\
\;\;\,i\\
        \end{array}\right), \\ O_5=\left(\begin{array}{l}
\;\;\,1\\
e^{\frac{\pi i}{4}}\\
\;\;\,i\\
e^{\frac{3\pi i}{4}}\\
-1\\
e^{\frac{5\pi i}{4}}\\
-i\\
e^{\frac{7\pi i}{4}}\\
        \end{array}\right), &  O_6=\left(\begin{array}{l}
\;\;\,1\\
e^{\frac{5\pi i}{4}}\\
\;\;\,i\\
e^{\frac{7\pi i}{4}}\\
-1\\
e^{\frac{\pi i}{4}}\\
-i\\
e^{\frac{7\pi i}{4}}\\
        \end{array}\right), & O_7=\left(\begin{array}{l}
\;\;\,1\\
e^{\frac{7\pi i}{4}}\\
-i\\
e^{\frac{5\pi i}{4}}\\
-1\\
e^{\frac{3\pi i}{4}}\\
\;\;\,i\\
e^{\frac{\pi i}{4}}\\
        \end{array}\right), & O_8=\left(\begin{array}{l}
\;\;\,1\\
e^{\frac{3\pi i}{4}}\\
-i\\
e^{\frac{\pi i}{4}}\\
-1\\
e^{\frac{7\pi i}{4}}\\
\;\;\,i\\
e^{\frac{5\pi i}{4}}\\
        \end{array}\right).
\end{array}   \label{X_eigenstate}
\end{equation}

\section{Detailed experimental statistical results.}\label{result}
In our experimental setup, the subspace is encoded in a rectangle within a plane using path-polarization DOFs to prepare an 8-dimensional entangled state:
\begin{equation}
    |\Phi\rangle=\sum_{k=0}^{7}|kk\rangle_{AB}, \label{quantum state}
\end{equation}
where the two-dimensional subspace of the prepared entangled state, such as $\{|0\rangle, |1\rangle\}$, was measured using the computational basis $\{|0\rangle, |1\rangle\}$ and the Fourier basis $\{(|0\rangle+|1\rangle)/\sqrt{2}, (|0\rangle-|1\rangle)/\sqrt{2}\}$ , yielding visibility of $0.999\pm0.001$ and $0.990\pm0.001$, respectively.\\ 

To precisely control the phase of the subspace and obtain the target quantum state (\ref{quantum state}), we modulate the phase between the following pairs of subspaces: $|0\rangle$ and $|4\rangle$, $|2\rangle$ and $|6\rangle$, $|1\rangle$ and $|5\rangle$, $|3\rangle$ and $|7\rangle$. This is achieved by adjusting the grayscale value of the corresponding regions in the SLM. Furthermore, the phases between $|0\rangle$ and $|2\rangle$, $|1\rangle$ and $|3\rangle$ are controlled by tilting BD1 and HWP1, while the phase between $|6\rangle$ and $|7\rangle$ is adjusted by tilting BD2.\\

After preparing the quantum state (\ref{quantum state}), we further modulate the grayscale in each SLM region and apply specific phases to designated subspaces to implement clock operation $Z^{x_1}$. For example, to add a phase of $i$ to the subspace $|1\rangle$, we construct projection measurements $(|0\rangle+|1\rangle)/\sqrt{2}$ and $(|0\rangle+i|1\rangle)/\sqrt{2}$ at the measurement ports of the two photons, adjusting the grayscale values of the SLM to minimize the joint probability. The fundamental idea is to design projection measurements that yield a joint probability of zero for entangled states within a two-dimensional subspace, based on the desired phase. Consequently, the clock operation $Z^{x_1}$ is successfully performed on Alice's photon through the SLM.\\

Due to the presence of a completely multi-outcome measurement device, it is crucial to ensure the coupling efficiency of each coupler for every subspace of the entangled state. In the quantum source generation, the BDs split the 404 nm laser and initiate a spontaneous parametric down-conversion (SPDC) process to generate 808 nm entangled photon pairs. The BDs in the measurement section overlap the 808 nm photons from different paths, forming a parallelogram configuration, such as the alignment between BD4 and BD5. By carefully adjusting the couplers and matching the BDs, the relative ratio of the coupling strength for each subspace of the prepared entangled state to each coupler can be maintained between 0.5 and 1. Despite the high visibility of the Fourier basis, reaching up to 0.99, the imbalance of the coupling strength of each significantly reduces the overall interference visibility.\\

In the experiment, for the prepared 8-dimensional maximally entangled state (\ref{quantum state}), we first select $x_1, x_2 = 00$, which corresponds to performing the operations $X^0$ and $Z^0$ on Alice's photons. This can be achieved through the method described above. Once Alice sends the encoded photons to Bob, he chooses the measurements $Z$ or $X$ to perform on the two photons based on the input $y \in \{1,2\}$.\\

For $y=1$, Bob sets the HWPs and QWPs in Fig.~\ref{measurementfig} to the angles listed in the first column of Table~\ref{angle}, thereby constructing the measurement $Z\otimes Z$  for the two photons. By statistically counting the photon coincidence events, we obtained the following success rate:
\begin{align}
    p(b=x_1 \mid x_1=0, x_2=0, y=1) = 0.9945. \notag
\end{align}

For $y=2$, Bob sets the HWPs and QWPs in Fig.~\ref{measurementfig} to the angles in the second column of Table~\ref{angle}, thereby constructing the measurement $X\otimes X$ for the two photons. By statistically counting the photon coincidence events, we obtained the following success rate:
\begin{align}
    p(b=x_2 \mid x_1=0, x_2=0, y=2) = 0.9496. \notag
\end{align}

Subsequently, we traverse $x = x_1 x_2$, where $x_1, x_2 \in \{0,1,\dots,7\}$, and perform the corresponding operation $U = X^{x_1} Z^{x_2}$ on Alice's photons. After Alice sends the encoded photons to Bob, he selects either the $Z\otimes Z$ or $X\otimes X$ measurement to perform on the two photons based on the input $y \in \{1,2\}$. By counting the photon-coincidence events, we obtained the remaining success rate $p(b=x_y \mid x_1, x_2, y)$. The experimental results are presented in Tables~\ref{result_Z} and \ref{result_X}. Therefore, in the entanglement-assisted quantum communication protocol for Schmidt number detection, the total success rate is
\begin{equation}
    \mathcal{S} = \frac{1}{2 n^2} \sum_{x_1, x_2=0}^{n-1} \sum_{y=1}^{2} p\left(b=x_y \mid x, y\right) = 0.9729,
\end{equation}
with $\sigma = 1 \times 10^{-4}$, where the standard deviation is estimated from Poisson statistics of photon counting through 1000 Monte-Carlo simulations.

\begin{table}[h]
    \centering
    \renewcommand\arraystretch{1.3}
    \begin{tabular}{|l|c|c|c|c|c|c|c|c|}
    \hline
       \diagbox{$x_1$}{$x_2$} & 0 & 1 & 2 & 3 & 4 & 5 & 6 & 7 \\
       \hline
       \quad0 & \;\;0.9945\;\; & \;\;0.9949\;\; & \;\;0.9942\;\; & \;\;0.9940\;\; & \;\;0.9942\;\; & \;\;0.9944\;\; & \;\;0.9941\;\; & \;\;0.9941\;\; \\
       \hline
       \quad1 & 0.9946 & 0.9950 & 0.9948 & 0.9945 & 0.9948 & 0.9950 & 0.9950 & 0.9948 \\
       \hline
       \quad2 & 0.9945 & 0.9944 & 0.9944 & 0.9948 & 0.9940 & 0.9942 & 0.9944 & 0.9945 \\
       \hline
       \quad3 & 0.9934 & 0.9937 & 0.9933 & 0.9934 & 0.9934 & 0.9935 & 0.9935 & 0.9933 \\
       \hline
       \quad4 & 0.9941 & 0.9942 & 0.9942 & 0.9940 & 0.9943 & 0.9944 & 0.9941 & 0.9940 \\
       \hline
       \quad5 & 0.9928 & 0.9926 & 0.9931 & 0.9929 & 0.9931 & 0.9932 & 0.9927 & 0.9928 \\
       \hline
       \quad6 & 0.9946 & 0.9946 & 0.9946 & 0.9947 & 0.9950 & 0.9944 & 0.9948 & 0.9947 \\
       \hline
       \quad7 & 0.9931 & 0.9929 & 0.9924 & 0.9931 & 0.9935 & 0.9930 & 0.9930 & 0.9931 \\
       \hline
    \end{tabular}
    \caption{After Alice applies the operation $U = X^{x_1} Z^{x_2}$ on her photons based on the input $x_1, x_2 \in \{0,1,\dots,7\}$, Bob performs a $Z \otimes Z$ measurement on the two photons, given the input $y=0$, to obtain the probability of winning $p(b=x_1 \mid x_1, x_2, y=1)$.}
    \label{result_Z}
\end{table}

\begin{table}[h]
    \centering
    \renewcommand\arraystretch{1.3}
    \begin{tabular}{|l|c|c|c|c|c|c|c|c|}
    \hline
       \diagbox{$x_1$}{$x_2$} & 0 & 1 & 2 & 3 & 4 & 5 & 6 & 7 \\
       \hline
       \quad0 & \;\;0.9496\;\; & \;\;0.9561\;\; & \;\;0.9507\;\; & \;\;0.9456\;\; & \;\;0.9541\;\; & \;\;0.9527\;\; & \;\;0.9515\;\; & \;\;0.9472\;\; \\
       \hline
       \quad1 & 0.9619 & 0.9487 & 0.9469 & 0.9423 & 0.9590 & 0.9473 & 0.9489 & 0.9407 \\
       \hline
       \quad2 & 0.9578 & 0.9618 & 0.9548 & 0.9644 & 0.9563 & 0.9621 & 0.9529 & 0.9618 \\
       \hline
       \quad3 & 0.9630 & 0.9486 & 0.9470 & 0.9409 & 0.9611 & 0.9463 & 0.9506 & 0.9425 \\
       \hline
       \quad4 & 0.9491 & 0.9526 & 0.9491 & 0.9447 & 0.9543 & 0.9539 & 0.9520 & 0.9479 \\
       \hline
       \quad5 & 0.9630 & 0.9481 & 0.9480 & 0.9321 & 0.9609 & 0.9475 & 0.9490 & 0.9426 \\
       \hline
       \quad6 & 0.9517 & 0.9615 & 0.9537 & 0.9642 & 0.9562 & 0.9588 & 0.9526 & 0.9605 \\
       \hline
       \quad7 & 0.9615 & 0.9475 & 0.9467 & 0.9321 & 0.9614 & 0.9467 & 0.9510 & 0.9431 \\
       \hline
    \end{tabular}
    \caption{After Alice applies the operation $U = X^{x_1} Z^{x_2}$ on her photons based on the input $x_1, x_2 \in \{0,1,\dots,7\}$, Bob performs a $X \otimes X$ measurement on the two photons, given the input $y=2$, to obtain the probability of winning $p(b=x_2 \mid x_1, x_2, y=2)$.}
    \label{result_X}
\end{table}

\section{Statistical significance} \label{App:StatSig}

Following the approach of \cite{Gill2002}, we express the statistical significance of our experimental results based on a binary hypothesis test.  Consider the random variables
\begin{equation}
\hat{\mathcal{S}}_i \equiv \sum_{bxy} \delta_{x_y,b} \frac{\chi(b_i = b,(x_1x_2)_i=x_1x_2,y_i=y)}{p\qty((x_1x_2),y)} \,,
\end{equation}
where $(x_1x_2)_i$, $y_i$ and $b_i$ denote respectively the setting of Alice, the setting of Bob and the outcome of Bob in the $i$th experimental run, $\chi(e)$ is the indicator function for the event $e$ (i.e., $\chi(e)=1$ if the event $e$ occurred while $\chi(e)=1$ if the event did not occur) and $p\qty((x_1x_2,y))$ denotes the joint prior on the settings of the devices, which were chosen uniformly, $p\qty((x_1x_2),y) = \frac{1}{2\times 8 \times 8}$.

The random variable $\hat{\mathcal{S}}_i$ is allowed to depend on previous experimental rounds, $j<i$, but not on future rounds $j>i$. We then define our estimator for $\mathcal{S}$ as the average of these random variables,
\begin{equation}
    \hat{\mathcal{S}} \equiv \frac{1}{N} \sum_i^N \hat{\mathcal{S}}_i \,,
\end{equation}
where $N \sim 2 \times 8 \times 8 \times 10^5 \sim 10^7 $ denotes the total number of experimental runs.

We express the $p$-value of our experiment, that is, the probability that the experimentally estimated success probability could arise from an entangled state of Schmidt number \emph{strictly less than eight} as \cite{Azuma1967},
\begin{equation}
p \qty(\hat{\mathcal{S}} \geq \mathcal{S}_{d=7} + \mu ) \leq \exp \qty( - \frac{2N \mu^2 }{\Delta^2}) \,,
\end{equation}
where $\mathcal{S}_{d=7}$ is the maximal attainable value of $\mathcal{S}$ with entangled states up to Schmidt number seven, $\mu \sim 5 \times 10^{-3} $ is the violation of that bound in our experiment and $\Delta \equiv \mathcal{S}_{\max} - \mathcal{S}_{\min} \leq 1$. Putting in the numbers, we find a vanishingly small $p$ value.

\printbibliography